\documentclass[useAMS]{emulateapj} 
\usepackage{natbib}
\usepackage{amsmath}
\usepackage{epsfig} 
\usepackage{times}

\def\be{\begin{eqnarray}}   \def\ee{\end{eqnarray}}
\def\ben{\begin{eqnarray*}} \def\een{\end{eqnarray*}} 

\def\sec#1{Section~\ref{sec:#1}}
\def\fig#1{Figure~\ref{fig:#1}} 
\def\tab#1{Table~\ref{tab:#1}}
\def\equ#1{Equation~(\ref{equ:#1})}

\newcommand{\Gyr}{\>{\rm Gyr}} 
\newcommand{\kpc}{\>{\rm kpc}} 
\newcommand{\ikpc}{\>{\rm kpc^{-1}}} 

\newcommand{\kms}{\>{\rm km}\,{\rm s}^{-1}} 
\newcommand{\kmstwo}{\>{\rm km}^{2}\,{\rm s}^{-2}} 
\newcommand{\Msun}{\>{M_{\odot}}} 
\newcommand{\Lsun}{\>{L_{\odot}}} 
\newcommand{\degree}{\ensuremath{^\circ}}
\begin{document}
\title {Galaxia: a code to generate a synthetic survey of the Milky Way}
\author{Sanjib Sharma, Joss Bland-Hawthorn\altaffilmark{1}}
\altaffiltext{1}{Leverhulme Visiting Professor, and Merton College Fellow, University of Oxford, OX1 3RH, UK.}
\affil{Sydney Institute for Astronomy, School of Physics, University of Sydney, NSW 2006, Australia}
\author{Kathryn V Johnston}
\affil{Department of Astronomy, Columbia University, New York, NY-10027} 
\author{James Binney}
\affil{Rudolf Peierls Centre for Theoretical Physics, 1 Keble Rd, Oxford, OX1 3NP, UK} 
\begin{abstract}
We present here a fast code for creating a synthetic survey of the
Milky Way. Given one or more color-magnitude bounds, a survey size 
and geometry,  the code
returns a catalog of stars in accordance with a given model of the
Milky Way. The model can be specified by a set of density 
distributions or as an N-body realization. We provide fast and
efficient algorithms for sampling both types of models. 
As compared to earlier sampling schemes which generate stars 
at specified locations along a line of sight, our scheme can generate 
a continuous and smooth distribution of stars over any given 
volume. The code is quite general and flexible and can accept  
input in the form of a star formation rate, age metallicity relation, 
age velocity dispersion relation and analytic density distribution
functions.  Theoretical isochrones are then used to generate a catalog
of stars and support is available for a wide range of photometric bands. 
As a concrete example we implement the Besan\c{c}on Milky Way
model for the disc.  For the stellar halo we employ
the simulated stellar halo N-body models of Bullock \& Johnston (2005). 
In order to sample N-body models, we present a scheme that disperses the stars
spawned by an N-body particle, in such a way that the phase space 
density of the spawned stars is consistent with that of the N-body particles.
The code is ideally suited to generating synthetic data sets that
mimic near future wide area surveys such as GAIA, LSST and HERMES. 
As an application we study the
prospect of identifying structures in the stellar halo with a simulated 
GAIA survey. We plan to make the code publicly available at 
{\tt http://galaxia.sourceforge.net}. 
\end{abstract}

\section{Introduction} 
Generating a synthetic catalog of stars 
in accordance with a given model of galaxy formation has a number of
uses. First, it helps to interpret the observational data. Secondly,
it can be used to test the theories upon which the models are
based. Moreover synthetic catalogs can be used 
to test the capabilities of different instruments, check for
systematics and device strategies to reduce measurement errors. 
This is well understood by the architects of galaxy redshift surveys
who rely heavily on $\Lambda$CDM simulations to remove artifacts imposed
by the observing strategy \citep{2001MNRAS.328.1039C}.

Given the widespread use of synthetic catalogs, a need for faster 
and accurate methods to generate stellar synthetic catalogs has recently 
arisen due to the advent of large scale surveys in astronomy, e.g.,
future surveys like LSST and GAIA have plans to measure over 
1 billion stars. In order to generate a synthetic catalog, one first
needs to have a model of the Milky Way. While we are far from a dynamically
consistent model, a working framework is fundamental to
progress. Inevitably, this will require approximations or assumptions
that may not be mutually consistent. Cosmologists already accept 
such compromises when they relate the observed galaxies to the dark-matter test 
particles that emerge from cosmological simulations.

There have been various attempts over the past few
decades to create a Galaxy model that is constrained by observations. 
The earliest such attempt was by 
\citet{1980ApJ...238L..17B,1980ApJS...44...73B,1984ApJS...55...67B}
where they assumed an exponential disc with magnitude dependent 
scale heights. An evolutionary model using population synthesis
techniques was presented by \citet{1986A&A...157...71R}. Given 
a star formation rate (SFR) and an initial mass function (IMF), one
calculates the resulting stellar populations using theoretical  
evolutionary tracks. Local observations were then used to constrain the 
SFR and IMF. \citet{1987A&A...180...94B} later introduced dynamical 
self consistency to constrain the disc scale height.
The present state of the art is described in  
\citet{2003A&A...409..523R} and is known as the Besan\c{c}on model.
Here the disc is constructed from a set of isothermal 
populations that are assumed to be in equilibrium. Analytic functions
for density
distributions, the age/metallicity relation and the IMF are provided for 
each population. A similar scheme is also used by the 
photometric code TRILEGAL \citep{2005A&A...436..895G}. 

In spite of its popularity, the current Besan\c{c}on model has important
shortcomings. A web interface exists to generate synthetic catalogs 
from the model but it has limited applicability for generating wide 
area surveys. Discrete step sizes for radial, and angular coordinates 
need to be specified by the user and results might differ 
depending upon the chosen step size. 
The scale height and the velocity dispersion of the disc are 
in reality a function of age but, due to computational 
complexity, the disc is modeled as a finite set of isothermal discs of 
different ages. Increasing the number of discs enhances 
the smoothness of the model but at the price of computational cost 
\citep{2005A&A...436..895G}.

In addition to the disc components, one also needs a model of the stellar halo. 
Under the hierarchical structure formation paradigm, 
a significant fraction of the stellar halo is thought to have been produced by 
accretion events and signatures of these should be visible as 
substructures in the stellar halo. Missions like GAIA, LSST and PanSTARRS 
are being planned that will enable us to detect substructures in the 
stellar halo.  

A smooth analytic stellar halo as in the Besan\c{c}on 
model is inadequate for testing schemes of substructure detection.
Furthermore, such a halo does not accommodate known structures
like the Sagittarius dwarf stream which may constitute a large
fraction of the present halo \citep{1995MNRAS.277..781I,2010ApJ...708.1290C}.
Substructures have complex shapes and hence to model them 
we cannot use the approach of analytic density distributions as
discussed earlier. However, N-body models are ideally suited for this task.
\cite{Brown2005} attempted to combine a smooth galaxy model with 
some simulated N-body models of disrupting satellites, but the 
stellar halo was not simulated in a proper cosmological context. 
Using hybrid N-body techniques, \citet{2005ApJ...635..931B}
have produced high resolution N body models of the stellar halo that are 
simulated within a cosmological context; see
also \citet{2010MNRAS.406..744C} and \citet{2008MNRAS.391...14D} for
a similar approach.
These can be used to make accurate predictions of the substructures 
in the stellar halo and also test the $\Lambda$CDM paradigm. 
However, as highlighted by \cite{Brown2005} there are several 
unresolved issues related to sampling of an N-body model and this has 
prevented their widespread use. 

The aim of this paper is to present fast and accurate methods 
to convert analytic and N-body models of a galaxy into a synthetic catalog
of stars. This would relieve the burden of generating catalogs from
modelers on one hand and on the other hand would allow the testing of 
models generated by different groups. 
This should also facilitate rapid testing of new models. 

We present a new scheme for sampling the analytical models that  
enables us to generate continuous values of the variables like
position and age of stars. Instead of a set of discs at specified ages,
our methodology allows us to generate a  disc that is continuous in
age.  As a concrete example, 
we use the Besan\c{c}on analytical model for the disc. 
To model the disc kinematics more
accurately, we employ the \citet{1969ApJ...158..505S} distribution
function that describes the non-circular motion in the plane of the disc.
For the stellar halo we use the simulated N-body models 
of \cite{2005ApJ...635..931B} that can reproduce the substructure in
the halo.  We show a scheme for
sampling the N-body particles such that the sampled stars preserve the 
underlying phase space density of N-body particles.

\section{Methods} \label{sec:methods}
\subsection{Analytic framework for modeling the galaxy}
We first describe the analytic framework that is used by us for modeling
the Galaxy. The stellar content of the Galaxy is modeled as a set of distinct 
components, e.g., the thin disc, the thick disc, the stellar halo and the bulge.
The distribution function, i.e., 
the number density of stars
as a function of position (${\bf r}$), velocity (${\bf v}$), age ($\tau$), 
metallicity ($Z$), and mass ($m$) of stars for each component is
assumed to be specified a priori. This can be expressed in general as   
\begin{eqnarray}
f_j & = &f_j({\bf r,v},\tau,Z,m), 
\label{equ:dist_func1}
\end{eqnarray}
$j$ being the label of the component. 
An accurate form of \equ{dist_func1} that describes all the properties 
of the Galaxy and is self consistent is still an open question. 
However, over the past few decades considerable progress has been made 
to arrive at a working model using a few simple assumptions
\citep{1986A&A...157...71R,1987A&A...180...94B,1997A&A...320..428H,1997A&A...320..440H,2005A&A...436..895G,2003A&A...409..523R}. Our 
analytical framework is based upon these models and we describe this below.

For a given galactic component, let the stars be formed at a rate of $\Psi(\tau)$ and the mass
distribution of stars $\xi(m,\tau)$ (IMF) be a parameterized 
function of age $\tau$ only. Also, let the present day 
spatial distribution  of stars, $f_{\rm pos}({\bf r},\tau)$ be a 
function age only. Finally, assuming that the 
velocity distribution $f_{\rm vel}({\bf v,r},\tau)$ and the 
metallicity distribution, $f_{Z}(Z,{\bf r},\tau)$ are a function of age
and position only one arrives at a model of the form 
(label $j$ omitted for brevity) 
\begin{eqnarray}
f_j & = &\frac{\Psi(\tau)}{\langle m\rangle}\xi(m,\tau)f_{\rm pos}({\bf r},\tau)f_{\rm vel}({\bf v,r},\tau)f_{Z}(Z,{\bf r},\tau)
\label{equ:dist_func2}
\end{eqnarray}
Note, the functions on the right hand side can potentially take
different forms for different galactic components, e.g., the thin disc,
the thick disc and so on.
The IMF here is normalized such 
that $\int_{m_{\rm min}}^{m_{\rm max}}\xi(m,\tau)dm =1$ and   
$\langle m \rangle=\int_{m_{\rm min}}^{m_{\rm max}}m\xi(m)dm$ is
the mean stellar mass.
We now discuss the functional forms of the metallicity and the velocity 
distribution.

The  distribution $f_{Z}$ is modeled as a log-normal distribution, 
\begin{equation}
f_Z=\frac{1}{\sigma_{\log\bar{Z}}(\tau)\sqrt{2\pi}}e^{-(\log Z-\log\bar{Z}(\tau))/(2\sigma_{\log
    \bar{Z}}^2(\tau))},
\label{equ:amr}
\end{equation}
the mean and dispersion of which is given
by an age-dependent function. The mean metallicity as function of age, 
$\bar{Z}(\tau)$, is popularly known as the age-metallicity relation (AMR).
In general a spatial dependence might also be added to \equ{amr},
e.g., $d\log(\bar{Z})/dR=$constant.

The velocity distribution $f_{\rm vel}$ can be modeled as a triaxial
Gaussian, 
\begin{eqnarray}
f_{\rm vel} & = &
\frac{1}{\sigma_{R}\sigma_{\phi}\sigma_{z}(2\pi)^{3/2}}  {\rm
  exp}\left(-\frac{v_{R}^2}{2\sigma_{R}^2(\tau,R)}-\frac{v_z^2}{2\sigma_z^2(\tau,R)}\right) \times
\nonumber \\
& & {\rm exp}\left(-\frac{(v_{\phi}-v_{{\rm
        circ}}(R)-v_{\rm ad}(\tau,R) )^2}{2\sigma^2_{\phi}(\tau,R)} \right)
\label{equ:veldist}
\end{eqnarray}
where, $R,\phi,z$ are the cylindrical coordinates,  
$v_{\rm circ}(R)$ the circular velocity as a function of
cylindrical radius $R$ and $v_{\rm ad}$ the asymmetric drift, which is
given by the Stromberg's relation \citep[equation 4.228 in][]{2008gady.book.....B} 
\be
v_{\rm ad}(\tau,R) & = & \frac{\sigma_{R}^2}{2 v_c} \times \nonumber \\
 & & \left( \frac{d \ln \rho}{d \ln
  R}+ \frac{d \ln \sigma_{\phi}^2}{d \ln R}+1-\frac{\sigma_{\phi}^2}{\sigma_R^2}+1-\frac{\sigma_z^2}{\sigma_R^2}\right)
\label{equ:stromberg}
\ee
Alternatively, one can model the distribution of 
$v_{\phi}$ using the \cite{1969ApJ...158..505S} distribution function
\citep{2009MNRAS.396..203S}, this is described in Appendix \ref{sec:rotational_kin}.
The dispersions of the $R,z$ and $\phi$ components of velocity 
increase as a function age due to secular heating in the disc 
and moreover there is also  radial dependence, the dispersion is
higher close to the center. We model these effects as in 
\citet{2010MNRAS.401.2318B} using the following functional form 
\begin{eqnarray}
\sigma_{R,\phi,z}^{\rm thin}(R,\tau) & = &\sigma_{R0,\phi0,z0}\left(\frac{\tau+\tau_{\rm
          min}}{\tau_{\rm max}+\tau_{\rm min}}\right)^{\beta} \left(\frac{\Sigma^{\rm thin}(R)}{\Sigma_{\odot}^{\rm thin}}\right)^q,
\label{equ:veldisp}
\end{eqnarray}
where $\Sigma$ is the surface density of the disc \citep[age
dependence from][]{2009MNRAS.397.1286A}. 

The model as given by \equ{dist_func2} has some limitations. 
For example, \equ{dist_func2} by itself is not 
dynamically self consistent and this has to be imposed externally.
For a given $v_{\rm circ}(R), \Psi(\tau)$ and $\sigma_z(\tau)$ 
the vertical structure of the disc 
in the solar neighborhood can be made dynamically self consistent by
using the approach of \citet{1987A&A...180...94B} \citep[see
also][]{2010MNRAS.402..461J}. This leads to a constraint of the form 
$\epsilon=\epsilon(\tau)$, $\epsilon$ being the ellipticity of a 
homoeoid disc (or scale height $h_{z}=h_z(\tau)$ for a double exponential disc).

The second limitation of \equ{dist_func2} is that it 
does not have an explicit dependence on the birth radii 
of a star.
Recently semi-analytic models of the Galaxy that treat 
chemical evolution with radial mixing have been 
proposed by \citet{2009MNRAS.396..203S} where the properties 
of a star are linked to their birth radii. 
A possible way to accommodate such models is 
to introduce in \equ{dist_func2} an additional parameter $R_i$, 
the radius at which a star is born in cylindrical coordinates. 
The model distribution function is then given by 
\begin{eqnarray}
f & = &\frac{\Psi(\tau,R_i)}{\langle m\rangle}\xi(m,\tau) \times
\nonumber \\
& & f_{\rm pos}({\bf r},R_i,\tau)f_{\rm vel}({\bf v,r},R_i,\tau)f_{Z}(Z,R_i,\tau),
\label{equ:dist_func3}
\end{eqnarray}
$f_{\rm pos}({\bf r},R_i,\tau)$ being the present day spatial distribution of
stars that were born $\tau$ years ago at radius $R_i$.

\subsection{Sampling an analytic model: adaptive von Neumann rejection
  technique} \label{sec:sampling_analytic}
Having specified the analytical framework we 
now discuss the scheme to sample stars from a model specified within 
this framework. In the present paper we restrict ourselves to
models of the form  of \equ{dist_func2} but the scheme we discuss 
is equally applicable to extensions of the form of \equ{dist_func3}.

For a model given by \equ{dist_func2},
if  ${\bf r}$ and $\tau$ are given, sampling the
metallicity distribution and the velocity distribution is trivial. 
Similarly, sampling the mass of a star is also straightforward as it is a 
one-dimensional distribution. 
Once the age, metallicity and mass of a star are known,  
a synthetic library of isochrones can then be used to generate 
stellar parameters like color, magnitude, gravity and so on.
So the key task is to sample the position and age 
coordinates from a specified analytical function 
\begin{equation}
g({\bf r},\tau)=\frac{\Psi(\tau)f_{\rm pos}({\bf r},\tau)}{\langle m\rangle},
\label{equ:gfunc}
\end{equation} 
which is multidimensional.

A simple approach to sample points from an analytical 
multidimensional distribution as given by \equ{gfunc} is to use the von Neumann rejection
technique --- generate a star having random coordinates ${\bf
  r},\tau$; then generate a random number $x$ between 
0 and $g_{\rm max}$ and finally accept the star if $x<g({\bf
  r},\tau)$.  
This naive approach is computationally very expensive. 
A lot of computational effort is spent to sample the low density 
regions that require rejection of a lot of stars. 
Moreover, the computational time does not scale with output sample size, so 
to generate even a small specific sample of stars, e.g., stars lying
within a given color magnitude limits or stars lying in some specific
region of space, one has to generate all the stars in the galaxy.

We now discuss an efficient technique to 
sample a  specified multidimensional 
distribution. 
We first divide the entire domain into bins that extend perpendicular 
to the time axis. Then we subdivide these bins into smaller 
sub-domains, called leaf nodes, with a spatial oct-tree  
(each sub node having 1/8th the volume of its parent node). 
Having subdivided the system, the von Neumann 
rejection technique is now applied individually to each of the nodes.
This has two immediate advantages:
\begin{itemize}
\item Depending upon the given geometry of the 
survey, one can check if the boundaries of the node intersect with 
that of the survey and, if not, one can skip the 
generation of stars.
\item For accepting and rejecting stars it is now possible 
to set the maximum density $g_{\rm max}({\bf r,\tau,{\bf l}})$, ${\bf
  l}$ being a vector representing the length of the sides of the node, 
adaptively for each node. Since the variation in stellar density  
in any node will be limited, $g/g_{\rm max}$ will never be very small,  
with the consequence that the rejection sampling method will not 
be inefficient.
\end{itemize}

We still need to decide on an optimum truncation criterion to stop the 
splitting of nodes. An 
ideal truncation criterion is to have the least density variation
within a node but this would result in a lot of nodes with negligible
number of stars but an otherwise strong number density gradient. 
Instead, we adopt a  
criterion that strives to achieve a fixed number of stars in a node 
similar to the one used in adaptive mesh refinement schemes. 
For a total of $N_{\rm tot}$ stars, we set the resolution limit 
of a node as $N_{\rm tot}/10^6$, a choice that effectively generates 
about $10^6$ nodes.
However, calculating the number of stars  in a node involves an
integral of the number density over the volume of the node, which can
be computationally intensive when done for a  large number of nodes. 
Hence, for the node
splitting purpose, we compute the number of stars as
\be
N_{\rm node}^{\prime}=g_{\rm max}({\bf r,\tau,{\bf l}}) l_1 l_2 l_3 l_4
\ee
where $g_{\rm max}({\bf r,\tau,{\bf l}})$ is an analytic density 
function representing  the maximum density within a node and 
$l_1, l_2, l_3, l_4$ are the length of the sides of the node.
The final number of stars in a node, $N_{\rm node}$, is calculated by 
numerically integrating the
number density over the node with increasing spatial and temporal resolution 
till the number converges with an uncertainty less than 
$0.1 \sqrt{N_{\rm node}}$.
The above convergence criterion roughly corresponds to $1/10$th of 
the expected Poisson noise.

To summarize, our scheme for generating a synthetic catalog of stars 
using the adaptive von Neumann sampling algorithm is as follows. 
For each node,
we first determine $N^{\rm node}$, the number of stars that need to be
generated.
Next, to assign position and age to stars, 
we randomly select ${\bf r}$ and $\tau$ and 
then accept or reject them using the von Neumann scheme. Next, the mass 
$m$ for a star is sampled from the IMF $\xi(m)$ 
and $Z$ from $f_{Z}(Z,{\bf r},\tau)$. The nearest isochrone corresponding to 
($Z,\tau$) is selected and the stellar parameters 
corresponding to $m$ are determined. The velocities are
assigned according to the distribution function $f_{\rm vel}({\rm
  v,r},\tau)$, which is
generally in the form of an ellipsoid or can be subjected to more
advanced treatment, an issue that we revisit later.
Finally, we use a selection function to decide if a 
star enters the final catalog of the survey. The process of spawning 
stars is repeated for all nodes to generate the full catalog. 

During the star spawning process we do not take into account 
the effect of photometric errors or extinction. This is because 
each observational survey will have different photometric error laws, 
and it is not possible to formulate a general scheme for this. 
Also, there is no perfect 3d model for extinction. Although, 
we do provide a default extinction model (see \sec{extinction}), 
but a  user might be interested in trying out other  models too. 
However, photometric errors and extinction can be easily handled 
as a post processing step and this is the approach that we follow. 
To accomplish this, first, one has   
to generate a catalog with color magnitude ranges slightly larger 
than the desired ranges and then for each star add the expected extinction 
and subsequently the photometric errors. Finally, using these transformed
magnitudes one can select the stars lying within the desired 
color magnitude limits.

\subsection{Speeding up the star spawning process}
The process of spawning stars from a node, as discussed in the
previous section, can be further optimized and we discuss this below.
To being with,
the distribution of stellar masses is such that there are a lot of low mass 
stars but very few high mass stars. Low mass stars are also low in 
luminosity hence for a magnitude limited survey only nearby low-mass  
stars are visible. On the other hand, the volume explored increases as 
the cube of the distance. Hence for most of the nodes the low mass
stars do not enter the final catalog. We use this information 
to optimize our scheme --- for each node we first determine the lowest 
stellar mass $m_{\rm min}^{\rm node}$ that can generate a visible
star and then exclusively generate stars having mass above this
limit. 
The number of stars from a node that fall into a survey is then given by
\begin{eqnarray}
N_{\rm vis}^{\rm node}(m_{\rm min}^{\rm node},m_{\rm max}) &=& {\int_{m_{\rm min}^{\rm node}}^{m_{\rm max}}
  \xi(m){\rm d}m}\int_{\rm node}
  g({\bf r},\tau){\rm d^3 r d}\tau \nonumber \\
&=&N^{\rm node} \xi(>m_{\rm min}^{\rm node}) 
\end{eqnarray}
We use stochastic rounding to convert
$N_{\rm vis}^{\rm node}$  to an integer --- i.e., if the fractional part of $
N_{\rm vis}^{\rm node}$ is less than a  Poisson-distributed random number with a range
between 0 and 1, we increment the integral part by 1.

We now describe the procedure to calculate $m_{\rm min}^{\rm node}$. 
If $r$ is the heliocentric distance of a node, having sides of length
$l_{\rm node}$, 
then calculating $m_{\rm min}^{\rm node}$ requires  
computing the minimum mass of a star that 
will be visible at a distance $r-\sqrt{3}l_{\rm node}$ for each isochrone and then 
computing the global minimum, i.e., over isochrones of all ages and 
metallicities. Doing this individually for each node might be 
computationally expensive. However, it is possible to optimize this. 
First, for a given galactic component, 
we identify the set of isochrones that have a
non-zero probability of being sampled and compute $m_{\rm min}^{\rm
  node}$ over these set of isochrones only. 
Secondly, we note that for a given set of isochrones  
 $m_{\rm min}^{\rm node}$ is a monotonically increasing function 
of heliocentric distance.  
Since, we can choose any value of $m_{\rm min}^{\rm
  node}$ that is smaller than the smallest visible mass,   
we access the nodes in a sequence sorted by their 
minimum heliocentric distance,  $r-\sqrt{3}l_{\rm node}$, 
and recompute $m_{\rm min}^{\rm node}$  whenever the distance 
changes by a fixed step size (a step size of 0.1 mag in distance modulus).

\begin{table*}
\caption{\label{tab:tb1} 
Geometry of  stellar components. The formulas used are from
\citet{2003A&A...409..523R}. Note, $(R,\theta,z)$ are the
  coordinates in the galactocentric cylindrical coordinate system and
  $a^2=R^2+\frac{z-z_{\rm warp}}{{\rm k_{\rm flare}}\epsilon(\tau)}$ (for the
  thin disc). } 
\begin{tabular}{lllll} \hline\hline 
Component & Age (Gyr)&density law $\rho({\bf r},\tau)$ \\ 
\hline 
Thin Disc	&$\le 0.15$ &$\frac{\rho_c \Psi(\tau)}{k_{\rm flare}\epsilon(\tau)} 
\{\exp(-(a/h_{R+})^2)-\exp(-(a/h_{R-})^2)\}$ \\
& & where: \ \ $h_{R+}$ = 5000 pc, $h_{R-}$ = 3000 pc \\ 
&  &  IMF- $\xi(m) \propto m^{-1.6}$ for $m<1 \Msun$ and $\xi(m) \propto m^{-3.0}$ for $m>1 \Msun$ \\ \\
Thin Disc &$0.15$--$10$ &$\frac{\rho_c\Psi(\tau)}{k_{\rm flare}\epsilon(\tau)}
\{\exp(-(0.5^2+\frac{a^2}{h_{R+}^2})^)-\exp(-(0.5^2+\frac{a^2}{h_{R-}^2})^)\}$ \\
& & where: \ \ $h_{R+}$ = 2530 pc, $h_{R-}$ = 1320 pc, \\ 
&  &  IMF- $\xi(m) \propto m^{-1.6}$ for $m<1 \Msun$ and $\xi(m) \propto m^{-3.0}$ for $m>1 \Msun$ \\ \\
Thick disc &11 & if $|z|\leq x_l, \ \ $ $\rho_c\delta(\tau-11)\exp{(-\frac{R-R_\odot}{h_{R}})} \times
  (1-\frac{1/h_{z}}{x_l\times (2.+x_l/h_{z})}\times z^{2})$ \\
&  & if
$|z|>x_l, \ \ $ $\rho_c\delta(\tau-11)\exp{(-\frac{R-R_\odot}{h_{R}})} \times
\frac{\exp(x_l/h_z)}{1+x_l/2h_z}\exp({-\frac{|z|}{h_{z}}})$ \\
& & where: \ \ $h_R$ = 2500 pc, $h_z$ = 800 pc, $x_l=400$ pc\\ 
&  &  IMF- $\xi(m) \propto m^{-0.5}$  \\ \\
Spheroid & 14 &$\rho_c\delta(\tau-14)\left(\frac{{\rm Max}(a_{\rm
    c},a)}{R_\odot}\right)^{n_{\rm H}}$\\
& & where: \ \ $a^2= R^2+\frac{z^2}{\epsilon^2}$, \\ 
& & $a_{\rm c}= 500$ pc, $\epsilon=0.64$, $n_{\rm H}=-2.77$  \\ 
& &  IMF- $\xi(m) \propto m^{-0.5}$  \\ \\
Bulge & 10&  if $\sqrt{x^2+y^2} < R_{\rm
  c}, \ \ $ $\rho_c\delta(\tau-10) \exp (-0.5 r_{\rm s}^2)$\\
& &  if $\sqrt{x^2+y^2} > R_{\rm c}, \ \ $ $\rho_c\delta(\tau-10)\exp (-0.5 r_{\rm s}^2)\times \exp (-0.5 (\frac{\sqrt{x^2+y^2}
 -R_{\rm c}}{0.5})^2)$\\
& & where: \ \  
$r_s^2=\sqrt{[(\frac{x}{x_0})^2+(\frac{y}{y_0})^2]^2+(\frac{z}{z_0})^4}$, \\
&  &  $R_{\rm c}=2.54$, $x_0=1.59$, $y_0=z_0=0.424$, $\alpha=78.9\degree$, $\beta=3.5\degree$, $\gamma=91.3\degree$\\ 
&  &  IMF- $\xi(m) \propto m^{-2.35}$  \\ \\
ISM & &$\rho_c  \exp(-\frac{R-R_\odot}{h_{R}}) \times
  \exp(-\frac{|z|}{h_{z}})$\\ 
&	&  where: \ \ $h_{R}$ = 4500 pc, $h_{z}$ = 140  pc\\  \\
Dark halo & & {\large $\frac{\rho_{\rm c}}{(1.+(a/R_{\rm c})^{2})}$}\\ 
& & where: \ \ $R_{\rm c}=2697$ pc and $\rho_{\rm c}=0.1079$\\ 
\hline 
\end{tabular} 
\end{table*}

\subsection{Sampling an N-body model}\label{sec:nbody_samp}
An N-body model consists of a finite set of particles distributed in phase
space ({\bf r,v}). Let $f_\textrm{phase-space}^\textrm{N-body}({\bf
  r,v})$ be the phase space distribution function describing this. 
An N-body particle in general represents a collection of
stars rather than a single star. Let $m^\textrm{N-body}$  
be the mass of the stellar population corresponding to the
N-body particle and this represents the mass of stars 
that were ever formed (summed over all ages). 
Also let us assume the age, metallicity and mass distribution 
to be given by  $\Psi(\tau)$, $f_{Z}(Z,\tau)$, and $\xi(m)$ 
respectively. The distributions being normalized such that $\int
\Psi(\tau)f_Z(Z,\tau)\xi(m) dZ d\tau dm=1$.  
The distribution function of stars is then given by
\begin{eqnarray}
\lefteqn{f^{*}=f_\textrm{phase-space}^\textrm{N-body}({\bf
  r,v}) \times} \nonumber \\
& &  \Psi(\tau)f_{Z}(Z,\tau)\xi(m) m^\textrm{N-body}/\langle m\rangle 
\end{eqnarray}
where the simulations provide the spatial and kinematic behavior
represented by the phase space density term $f_\textrm{phase-space}^\textrm{N-body}({\bf
  r,v})$. 

For an N-body model, the fact that the number of N-body particles
is finite leads to the following problem --- if the
number density of visible stars at a given point in space 
is more than the number density of N-body particles at the same point
one needs to oversample the particles. The condition for this 
is given by
\begin{eqnarray}
\int d\tau dZ \int_{m_{\rm min}(r)}^{m_{\rm max}}  f^{*} dm & > & f_\textrm{phase-space}^\textrm{N-body}({\bf
  r,v}), \nonumber 
\end{eqnarray}
or in other words 
\begin{eqnarray}
\xi(>m_{\rm min}(r))\frac{m^\textrm{N-body}}{\langle m \rangle} & > & 1, 
\end{eqnarray}
$m_{\rm min}(r)$ being the
minimum mass of a star that is visible at  a heliocentric distance of $r$.
Typically such a situation arises when $r$  is small, i.e., in regions
close to the sun. 
Hence unless one simulates the N-body model with the number of particles
equal to the number of stars in the model galaxy, one has to live with oversampling. The
question now is how are we to assign positions and velocities to the
oversampled particles?

The positions of oversampled particles should be assigned in such a way
that they preserve the underlying phase space density. 
For this we first review a popular scheme of density estimation 
known as the kernel density estimation, which is also statistically 
quite robust. In this scheme 
the number density at a given point ${\bf x}$  for an N-body system of particles is given by
\begin{equation}
f({\bf x})=\sum_{j} \frac{1}{h_j^n}K\left(\frac{r_j}{h_j}\right)  
\end{equation}
where $K(u)$ is a kernel function such that its integral over all
space is unity, $r_{j}$ is the distance of the $j$-th particle
from ${\bf x}$, and $n$ the dimensionality of the space. In a multivariate setting, 
$r_j$  is given by 
\begin{equation}
r_j=\sqrt{\bf (x-x_j)^{T}\Sigma_j^{-1}(x-x_j)}
\end{equation}
where ${\bf \Sigma_j^{-1}}$ is the matrix representing the
distance metric and is normalized such that $ {\rm det}({\bf \Sigma_j})=1$.
The smoothing length $h$ of a particle is  determined as the 
distance to the $k$-th 
nearest neighbor from the location of the particle. 
The physical interpretation of the density scheme is that if each point is 
assumed to be a hyper-ellipsoidal ball with mass distributed according
to the kernel function $K$ and having a covariance of $h^2\Sigma$, 
then the underlying smooth density field is simply the result of 
superposition of such balls. 
This suggests that if the spawned stars are also distributed around 
the particle in a similar fashion, i.e., distributed according to 
the kernel function $K$ and having a covariance is $h^2\Sigma$, 
then they will essentially 
be sampling the density field of the N-body system. 

The question that remains is 
how to determine the optimum covariance matrix  $\Sigma$?
The simplest scheme is to use a global metric based on the 
variance of the data along each dimension.
A global metric, however, is
insufficient for accurately calculating the phase space densities.
What is needed is a locally adaptive metric that changes with the 
local configuration of the data at each point in space
\citep{2006MNRAS.373.1293S,2009ApJ...703.1061S}.
Such a metric can be calculated using the publicly available 
multidimensional density estimation code EnBiD by 
\citet{2006MNRAS.373.1293S}.
EnBiD, which is an improvement over 
a scheme proposed by \citet{2005MNRAS.356..872A}, makes use of a binary
space partitioning scheme along with the use of the information theoretic
concept of Shannon entropy to handle the issue of determining the
optimal locally adaptive distance metric $\Sigma_j^{-1}$ in a 
multi-dimensional space. 
For our use, we adopt the EnBiD scheme with the number of nearest neighbors 
$k$ set to 64, which we find is adequate for reproducing a smooth
distribution of stars in the phase space. Too small a value leads to 
lumpiness around the N-body particles and too large a value erodes 
the positional accuracy of the data. We assume the matrix
 $\Sigma$ to be diagonal and use the cubic cell scheme of EnBiD, 
i.e $\Sigma_{11}= \Sigma_{22}= \Sigma_{33}$ and $\Sigma_{44}=
\Sigma_{55}= \Sigma_{66}$ (dimensions 1,2 and 3 representing
position coordinates while the dimensions 4,5 and 6 the velocity 
coordinates). The EnBiD algorithm was applied individually to each 
of the disrupted satellites in the simulations. 

\begin{figure}
  \centering \includegraphics[width=0.50\textwidth]{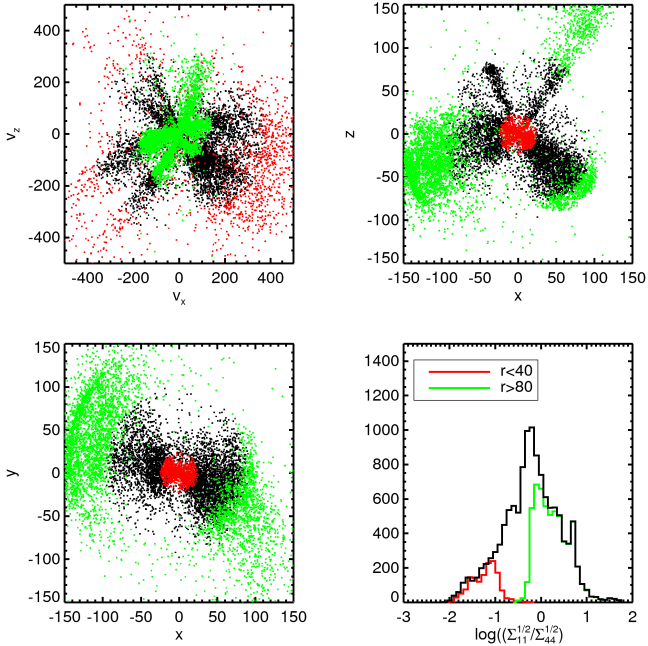}
\caption{Distribution of the velocity scale factor in units of 
$\kpc/\kms$ for  one of the disrupted satellites from BJ05 simulations and its 
 dependence on location in phase space. The bottom left panel 
shows the distribution of velocity scale factor 
 while the other panels show 
the scatter plot of particles in position and velocity space. The
particles lying in the inner ($r<40$) and outer regions ($r>80$) 
of the halo are colored red and green respectively 
($r$ being the radial distance from the center of the stellar halo).
High density regions in position space are low density regions in
velocity space and vice versa.
\label{fig:smoothing_length}}
\end{figure}

In order to assess the usefulness of EnBiD for our particular
application,  we plot in \fig{smoothing_length} (lower right panel) 
the distribution of
the velocity scale factor $\sqrt{\Sigma_{11}/\Sigma_{44}}$ , 
which is the ratio of position to velocity scale and has units of 
$\kpc/\kms$, as estimated 
by EnBiD for the N-body particles from one of the simulations of 
a disrupted satellites by \cite{2005ApJ...635..931B} 
(see \sec{simulated_halo}). 
It can be seen that the scale factor varies by about 3 orders of magnitude, 
which demonstrates the inappropriateness of a global metric 
and the need for a locally adaptive metric scheme. The huge variation 
is due to the fact that during phase mixing the phase space density 
of particles is approximately conserved. This means that high density
regions in position space will 
correspond to low density regions in velocity space and vice versa. 
This can be easily seen in the panels of  \fig{smoothing_length}, 
where scatter plots of particles in position and velocity space has
been shown. 
The high density regions in position space ($r<40$) are
colored red while the low density regions in position space 
($r>80$) are colored green. 
The mean value of scale factor $\langle \sqrt{\Sigma_{11}/\Sigma_{44}}
\rangle$ was found to be $0.057$ and $2.9$ for the red and green
regions (see lower right panel for distributions). As a check we 
computed the scale factor using variance of the particles
$\sqrt{(\sigma_{\rm x}^2+\sigma_{\rm y}^2+\sigma_{\rm z}^2)/(\sigma_{\rm vx}^2+\sigma_{\rm vy}^2+\sigma_{\rm vz}^2)}$.
The values obtained were  $0.037$ and $2.3$ for red and green particles
respectively, which are in agreement with results reported by EnBiD.

Summarizing, the scheme for sampling N-body distributions is as
follows. Select a particle $i$, determine if the dispersing volume, 
defined by the kernel function $K$ and the smoothing length $h_i$,
intersects with the survey volume, and calculate the number of visible
stars it spawns, i.e., $N_{\rm vis}^i=m^\textrm{N-body}\xi(>m_{\rm min}(r))/\langle m
\rangle$. We use stochastic rounding as discussed earlier to convert
$N_{\rm vis}^i$  to an integer.
Next, for each spawned star we generate $\tau$, $Z$ and $m$ 
according to the distribution functions $\Psi_{\tau}$, $f_{Z}(Z,\tau)$ and
$\xi(m)$ and disperse them in phase space using the 
scheme discussed above.
Finally, we use a selection function to decide if a 
star enters the final catalog of the survey.

\begin{figure}
  \centering \includegraphics[width=0.50\textwidth]{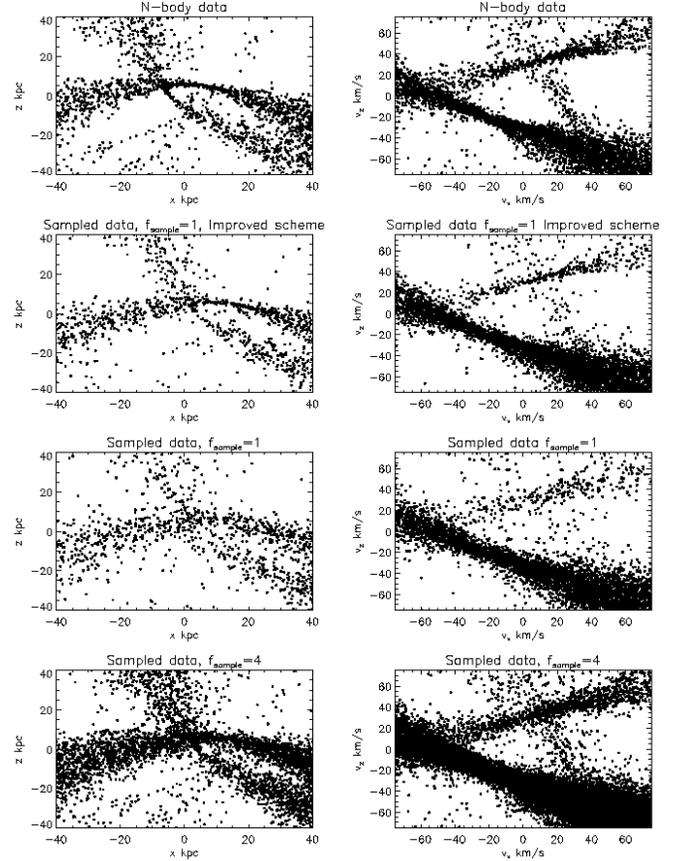}
\caption{ Position and velocity scatter plots of a disrupted satellite 
sampled from an N-body simulation of \cite{2005ApJ...635..931B}. 
The upper panels show the original N
body data while the rest of them are for sampled data. The sampling
fraction $f_{\rm sample}$ is labeled on the panels. The panels in second 
row use an improved scheme that disperses the spawned stars in phase
space only when more than one star is spawned by an N-body particle.
Note, the stellar mass associated with an N-body particle is not
necessarily constant. 
\bigskip
\label{fig:sampling_test}}
\end{figure}

\subsubsection{Improving the positional accuracy 
while sampling an N-body model}
Dispersing the stars spawned by an N-body particle 
in phase space has its side effects. This adds noise to the data, the 
effect of which is to reduce the positional accuracy. 
When  a particle is oversampled  one cannot avoid the noise 
but when a particle is under-sampled or equally-sampled i.e, less than or 
equal to one star needs to be spawned from the N-body particle, one 
can eliminate the noise by simply assigning the coordinates of the
particle directly to the spawned star. 
But it is not possible to know a priori as to how many stars 
from a particle will enter final catalog. To overcome this we
implement a scheme which makes sure that, among the stars spawned
by a particle if one of the stars can be reassigned the coordinates of 
its parent particle without it being excluded from the final catalog 
we do so. The effect of this improved sampling
scheme is shown in \fig{sampling_test} which shows the position (left panels)
and velocity (right panels) scatter plot of a disrupted satellite system from an
N-body simulation of \cite{2005ApJ...635..931B}. The panels in the top row  
show the N-body particles in the simulation, while the other rows
show the stars sampled from it. The sampling fraction $f_{\rm sample}$,  
i.e., average number of spawned stars per N-body particle, is also
labeled on each of the panels. For illustrative purpose we assume
the number of stars spawned per unit mass of the particle 
to be constant but in a realistic simulation it will 
depend upon the heliocentric distance of the particle.
First, we compare the first row 
with the second row that shows the results of the improved scheme 
for the case of $f_{\rm sample}=1$. 
The distribution of position coordinates is nearly identical 
for both rows which is what we expect when $f_{\rm sample}=1$. 
On closer scrutiny, in some regions a loss of positional accuracy 
can be seen. Also some regions apparently seem to be undersampled. 
This departure from the ideal behavior is because the stellar mass 
associated with an N-body particle obtained from the simulations 
that we use, is not necessarily constant. So some N-body particles can 
spawn more than one star which subsequently needs to be dispersed in 
phase space.
Next, we compare
the results of the improved scheme (second row) with the naive scheme 
(third row) that disperses all stars for the same value of $f_{\rm sample}=1$. 
It can be clearly seen that the improved scheme has better 
positional accuracy. For reference, the last row shows the case 
of $f_{\rm sample}=4$. We revisit this issue in \sec{gaia_anal} and 
\sec{rvr_anal} and discuss in greater detail the consequences of 
our scattering scheme for real applications.

\subsection{Stellar Isochrones}
Theoretical stellar isochrones are one of the main ingredients of our 
scheme. These are used to assign parameters like 
luminosity, effective temperature, magnitude and color to a star 
of a given age, metallicity and mass. 
We employ Padova isochrones 
\citep{2008A&A...482..883M,2007A&A...469..239M,2000A&AS..141..371G,
1994AAS..106..275B}, which are available for a wide variety of 
photometric systems 
\footnote{The isochrones were downloaded from
{\tt http://stev.oapd.inaf.it/cgi-bin/cmd} }
and unlike other popular isochrones 
also cover the asymptotic giant branch and the red clump phases of 
stellar evolution.

We now describe the scheme for interpolating across different isochrones.
An isochrone, for a given age and metallicity, is a table of stellar 
parameters, e.g., magnitudes, effective temperature and gravity, 
as a function of stellar mass. Hence in a given isochrone, stellar parameters 
for any intermediate mass can be easily computed by linear interpolation.
However, interpolating across isochrones of different ages and 
metallicities involves the use of equivalent evolutionary
points, which are not always provided with the isochrones. 
Hence, for the sake of flexibility as well  as computational 
ease we adopt a simpler scheme. 
To begin with, a grid of $182 \times 34$ isochrones spanning an age 
interval of  
$6.6<\log{t/{\rm yr}}<10.22$ (step size of $\Delta \log(t)=0.02$ )
and metallicity interval of $10^{-4}<Z<3 \times 10^{-2}$ (mean step size of
$\Delta \log(Z) \sim 0.072$) were chosen. 
Next, for a star of a given age and metallicity, the 
nearest isochrone in the grid was identified.
Subsequently, this isochrone was used to
compute the stellar parameters by linear interpolation in mass.
Where needed, the metallicity $Z$ was converted to [Fe/H] using the
relation 
\begin{equation} 
[{\rm Fe/H}]=\log({Z/Z_{\odot}}); \ \ {\rm where} \ {Z}_{\odot}=0.019  
\end{equation}

In the above mentioned scheme, to simulate the color 
magnitude diagrams accurately, the adopted grid should have a fine
enough resolution; and 
our adopted resolution was found to be adequate for this purpose.  
However, for greater accuracy if desired one can increase the 
resolution of the grid. 

The Padova isochrones are limited to stellar masses greater
than $0.15 \Msun$. For lower masses extending up to the hydrogen mass 
burning limit ($0.07 \Msun < m < 0.15 \Msun$) we use the isochrones from 
\citet{2000ApJ...542..464C}. Since, these isochrones are only available
for solar metallicity,   we used them for all metallicities. 
Also, the \citet{2000ApJ...542..464C} isochrones are only available for 
Johnson-Cousin bands, hence, in order to able to support  
a variety of photometric bands, we  choose to compute absolute magnitudes 
for the above mentioned low mass stars using the bolometric correction 
(BC) tables from
{\tt http://stev.oapd.inaf.it/dustyAGB07/}, which is a database of 
BC tables for different photometric bands provided by the Padova group 
\citep{2008A&A...482..883M,2002A&A...391..195G,2004A&A...422..205G}. 

Note, in the present version of the code we do not model the 
white dwarfs. White dwarfs are quite faint ($M_V>10$) and
rare compared to other type of stars and hence for most 
applications they do not dominate the star counts. 
However, in applications where one expects to find white dwarfs 
caution should be exercised when interpreting the results 
of {\sl Galaxia}.

\subsection{Extinction}\label{sec:extinction}
Most stellar observations suffer from extinction by 
dust and this needs to be taken into account when 
comparing simulated catalogs with stellar surveys. 
Although the total extinction along a given line of sight has been 
accurately mapped \citep{1998ApJ...500..525S}, the 
same cannot be said for the 3D distribution of dust.  
\cite{2003A&A...409..205D} provide a sophisticated model 
for the 3D dust distribution, and other alternatives include 
\citet{2006A&A...453..635M}. 
Since a comprehensive extinction model 
is beyond the scope of this paper, we provide here only a simple extinction 
correction scheme. 
Note, extinction is basically a post processing step and once 
an extinction-free catalog has been generated using {\sl Galaxia}, one can 
subject it to the extinction model of one's choice.  

The extinction scheme we use is a refinement of the method proposed by
\citet{2010ApJ...713..166B}.
The 3D distribution of the  dust is modeled as a double exponential
disc,  
\be
\rho_{\rm Dust}(R,z) =\frac{\rho_0}{k_{\rm flare}} \exp\left(-\frac{R-R_{\odot}}{h_{R}}\right) \times
  \exp\left(-\frac{|z-z_{\rm warp}|}{k_{\rm flare}
      h_{z}}\right). \nonumber \\	 
\ee
where $z_{\rm warp}(R,\phi)$ and $k_{\rm flare}(R)$ describe the 
warp and flare of the disc and are described in \sec{warpflare}. 
The values of the parameters controlling the warp and flare were 
assumed to be same as that of the stellar disc. 
Next, we use the \citet{1998ApJ...500..525S} $E(B-V)$ dust maps to 
evaluate the best fit parameters of the
dust model and obtained $h_{R}$ = 4200 pc, $h_{z}$ = 88 pc and 
$\rho_0=0.54 {\rm mag} /\kpc$. 
The fit parameters were obtained by minimizing
\begin{equation}
\chi^2=\sum
\left(\frac{E_i^{\rm map}-E_i^{\rm model}}{E_i^{\rm map}}\right)^2
\end{equation}
where
$E_{i}^{\rm model}=\int_{0}^{\infty} \rho_{\rm Dust}(l_i,b_i,r) dr$ is the extinction in
the cell $i$ defined in the $(l,b)$ space. The fit was performed in
the range $-15\degree < b < 15\degree$ assuming $R_{\odot}=8.0$ which
defines the distance scale.

The modeled dust disc is then used to generate a
3D extinction map in galactic coordinates $l,b$ and $r$, 
by integrating the dust density along different lines of
sight.
The angular resolution 
of the 3D maps was assumed to be $1024\times1024$ while in the radial 
directions 512 logarithmically spaced bins in the range $0.01<r<30
\kpc$ were used. 
Values for any arbitrary position were obtained from the maps 
by linear interpolation in $(r,l,b)$. 
Alternately, one can also assume a directional dependent normalization 
constant $\rho_0$ such that the total extinction at infinity along a 
given line of sight equals the value in the \citet{1998ApJ...500..525S} maps. 
For this we use high resolution 
\citet{1998ApJ...500..525S} 
maps having an angular resolution of $4096\times4096$ in $l$ and $b$. 
We employ this as our default scheme.

Note, Schlegel extinction maps are known to overestimate the 
extinction in the Galactic plane (regions where $A_V>0.5$ mag), 
by about $30\%$ \citep{1999ApJ...512L.135A,2005A&A...435..131C}. 
While \cite{1999ApJ...512L.135A} report results in a strip of few 
degrees in length only, \cite{2005A&A...435..131C} report 
mean values in galactic anti-center hemisphere only. 
Since a proper recalibrated map is not yet 
available it is difficult to take these effects into account.
In addition to directly affecting the total value of extinction 
along a particular line of sight, the overestimation is also expected to 
affect the determination of our scale height for the dust disc, which 
in turn will affect the variation of extinction with distance.
If we assume the overestimation factor to be a monontonic function of 
extinction we expect the actual scale height of the dust disc 
to be larger than what we find here. 
Hence, we advise caution to be exercised when using our extinction 
scheme in the galactic plane.

\section{A working model of the galactic components}
The main ingredients of our galactic model as described by  \equ{dist_func2} 
are the star formation rate 
(SFR), the age velocity relation (AVR), the IMF and the density profiles.
We now discuss the adopted functional forms for each of these 
functions so as to arrive at a working model of the Galaxy.
Instead of doing a detailed modeling we here implement the well known 
and well tested Besan\c{c}on model.  
The density distributions and the ages and metallicity  of 
various galactic components are given in
\tab{tb1} and \tab{tb2}, these are the same as the one used by
\cite{2003A&A...409..523R} (R03 hereafter). 
The thin disc spans an age interval of
$0-10 \Gyr$. On the other hand, the thick disc, the bulge and the halo   
are all assumed to have a constant age. 

\begin{table}
\caption{\label{tab:tb2} Age and metallicity distribution (mean and 
dispersion) of  galactic components. The values shown are from \cite{2003A&A...409..523R}} 
\begin{tabular}{lllll} 
\hline
 & Age (Gyr) & $\rm [Fe/H]$ &$\sigma_{\rm [Fe/H]}$ & ${\rm d[Fe/H]/
   dR}$ \\
\hline
Thin disc & 0-0.15 & -0.01 & 0.12 &       \\
          & 0.15-1 & -0.03 & 0.12 &       \\
          & 1-2    & -0.03 & 0.10 &       \\
          & 2-3    & -0.01 & 0.11 & -0.07 \\
          & 3-5    & -0.07 & 0.18 &       \\
          & 5-7    & -0.14 & 0.17 &       \\
          & 7-10   & -0.37 & 0.20 &       \\
\hline
Thick disc & 11    & -0.78 & 0.30 &  0    \\
\hline
Stellar Halo & 14  &-1.78  & 0.50 &  0    \\
\hline
Bulge      & 10 & 0.00     & 0.40 &  0    \\
\hline
\end{tabular}
\end{table}

\subsection{Thin and thick disc}
In the Besan\c{c}on model, the axis ratio  $\epsilon$ and velocity
dispersions $\sigma_R,\sigma_{\phi}$ and $\sigma_{z}$, for the thin
disc, are tabulated 
as a function of age $\tau$ (velocity ellipsoid being taken from 
\citet{1997ESASP.402..621G}. 
Here we parameterize them as functions. The axis ratio $\epsilon$ 
is parameterized as
\be
\epsilon(\tau)={\rm Min}\left(0.0791,0.104\left(\frac{\tau+\tau_{\rm min}}{\tau_{\rm max}+\tau_{\rm min}}\right)^{0.5}
  \right),
\ee
and for velocity dispersions we use  \equ{veldisp}. 
To match the Besan\c{c}on values, the velocity dispersion 
for the thin disc is assumed to saturate at $0.862 \sigma_{R0,\phi
  0,z0}$ and a value of $\tau_{\rm max}=10.0$ and $\tau_{\rm min}=0.1$ 
is used. The values of  parameters used are summarized in \tab{tb3}. 
The adopted values for the thin and thick disc 
reproduce the  Besan\c{c}on results.
Note, the radial dependence of velocity dispersions 
is modeled here using \equ{veldisp} \citep[values of $q$ and $\beta$ 
being from][]{2009MNRAS.396..203S}, rather than the Besan\c{c}on model
that assumes $d\ln \sigma_R^2/dR=-0.2 \kpc^{-1}$ and zero derivative
for other components. 
Note, the circular velocity profile is computed from the 
Besan\c{c}on mass model and at the location of the Sun 
it has a value of $226.84 \kms$.

\begin{table}
\caption{\label{tab:tb3} 
Velocity ellipsoid of  stellar components. Note, $(R,\phi,z)$ are the
  coordinates in the galactocentric cylindrical coordinate system. } 
\begin{tabular}{llllll} \hline
& $\sigma_{R0}$ & $\sigma_{\phi0}$ & $\sigma_{z0}$& $q$& $\beta$ \\
& kms/s & kms/s & kms/s & &  \\
\hline
Thin disc  & 50  & 32.3 & 21 & 0.33  & 0.33 \\
Thick disc & 67  & 51   & 42 & 0.33  & 0.33  \\
Spheroid   & 141 & 75   & 75 & 0     & 0  \\
Bulge      & 110 & 110  & 100& 0     & 0  \\
\hline
\end{tabular}
\end{table}

\subsection{Warp and flare}
\label{sec:warpflare}
The thin and thick discs are assumed to have a warp and a flare that  
is modeled following
the prescription of R03. 
Assuming galactocentric cylindrical coordinates $(R,\phi,z)$, 
stars with radius $R>R_{\rm warp}$ are displaced perpendicular to the 
plane by an amount
\begin{equation}
z_{\rm warp}(R,\phi)=\gamma_{\rm warp} {\rm Min}(R_{\rm warp},R-R_{\rm
  warp}){\rm cos}(\phi-\phi_{\rm max})
\end{equation}
where $\phi_{\rm max}$ is the direction in which the warp is maximum. 
For flaring, stars with $R>R_{\rm flare}$ have their scale heights
increased by a factor
\begin{equation}
k_{\rm flare}(R)=1+\gamma_{\rm flare} {\rm Min}(R_{\rm flare},R-R_{\rm
  flare})
\end{equation}
For the parameters we adopt the same values as that used by 
R03, namely 
$\phi_{\rm max}=90.0$, $\gamma_{\rm warp}=0.18\ikpc$, $R_{\rm flare}=1.12R_{\odot}$
and $\gamma_{\rm flare}=0.0054\ikpc$.

\subsection{Bulge}
It has been well established that the Milky Way hosts a bar-shaped 
bulge \citep{1993AIPC..278...98B}. With the arrival of the COBE 
infrared maps, it has been possible to construct 3d models to fit the 
data. Various models using different triaxial analytic 
functions were presented by \citet{1995ApJ...445..716D}, of which 
the G2 model (boxy triaxial Gaussian type functions) provided the best
fit to the data.  The G2 bar is used by the Besan\c{c}on model and is
adopted by us in our initial demonstration of the {\sl Galaxia} model.
The G2 density distribution has a core at the center 
defined by radius $R_c$ and the distribution is elongated along 
one axis which defines the major axis. The orientation of the major 
axis is defined by three 
angles $\alpha, \beta$ and $\gamma$. The values for these parameters are 
taken from R03, where they obtained the values by fitting 
the model to the near infrared star counts of the DENIS survey.

Next, we describe the kinematic properties of our model. 
Self-consistent dynamical models have been presented that use either 
the Schwarzschild technique \cite{1996MNRAS.283..149Z} or are extracted
from N-body simulations \citep{1999A&A...345..787F,2002MNRAS.330...35A}, 
but these are beyond the
scope of the current paper. Instead, we attempt to create 
a simple working model that roughly satisfies the 
existing observational data and would be useful for studying 
the systematic effects that cloud the interpretation of observation data. 

A number of earlier studies have claimed that the bulge is in solid 
body rotation by fitting a straight line to the plot of mean radial 
velocity as a function of longitude (also referred to as a rotation curve)
, e.g., by \cite{1990ESOC...35..115M} using 
Mira variables and by \cite{1995ApJ...453..837I} using SiO maser stars. 
A slope of about $10 \kms{\rm deg^{-1}}$ has generally been reported,  
which translates to a pattern speed of $\Omega=71.62 \kms\ikpc$, 
assuming $8 \kpc$ as the distance to the galactic center.
Recently, \citet{2009ApJ...702L.153H} have also measured the rotation
curve using bulge M-giants in two strips at $b=-4$ and $b=-8$. They 
report the two rotation curves to be nearly identical, 
suggesting that the bulge is rotating cylindrically. 
A straight line fit (by eye) to their rotation curve 
also seems to suggest a value close to  $10 \kms{\rm deg^{-1}}$. 
Hence, for simplicity we adopt the value of $\Omega=71.62 \kms\ikpc$ 
for the pattern speed of the bulge. 
Note, \cite{2008ApJ...688.1060H} also report that the rotation 
curve  for the $b=-4$ strip, shows evidence of flattening 
beyond $|l|>4$. 

Having specified the rotation we next review the velocity 
dispersions. Specifically, we attempt to fix the values for 
the velocity dispersions $\sigma_R,\sigma_\phi$ and $\sigma_z$ 
of our model, expressed in cylindrical coordinates with respect 
to the galactic center.   
A value of $\sigma_r=110.0 \kms$ has typically been reported 
for the line of sight velocity dispersion in Baade's window 
\citep{1995AJ....110.1774T,1998gaas.book.....B} and this is what 
we assume for $\sigma_R$. Recent results 
by \cite{2009ApJ...702L.153H,2008ApJ...688.1060H} also confirm 
this but show a variation with both latitude and longitude-- 
at $b=-4$, $\sigma_r$ is found to increase from $70 \kms$ to 
$110 \kms$ with $|l|$ varying from $10\degree$ to $0\degree$, 
while at $b=-8$,  $\sigma_r$ is found to be relatively constant 
with a value of $70 \kms$. 

Using proper motions, velocity dispersions along $l$ and $b$ 
have also been measured and an anisotropy of 
$\sigma_l/\sigma_b =1.15$ has been reported
\citep{2007MNRAS.378.1165R,2007AJ....134.1432V,1992AJ....103..297S}. 
The anisotropy could both be due to intrinsic anisotropy as well as
due to the rotational broadening as demonstrated by 
\cite{1996ApJ...470..506Z}. Since it is not possible to exactly deduce 
the value of $\sigma_{\phi}$ from $\sigma_{l}$ we assume for
simplicity $\sigma_{\phi}=\sigma_{R}$ and choose the value of $\sigma_z$ 
so as to reproduce the observed value of anisotropy ratio 
in Baades's window. For a choice of $\sigma_z=100 \kms$, 
the bulge stars lying in the field $(l,b)=(1\degree,-4\degree)$ and 
with heliocentric distance between $7-9 \kpc$, were 
found to have  $\sigma_l/\sigma_b=1.17$.

\subsection{Simulated stellar halo} \label{sec:simulated_halo}
To make theoretical predictions of structures in the stellar halo, 
we use the eleven stellar
halo models of \cite{2005ApJ...635..931B}, which were simulated within the
context of the $\Lambda$CDM cosmological paradigm.
These simulations follow the accretion of individual satellites modeled as
$N$-body particle systems onto a galaxy whose disc, bulge, and halo potential
is represented by time dependent analytical functions. Semi-analytical
prescriptions are used to assign a star formation history to each satellite
and a leaky accreting box, chemical enrichment model is used to calculate the
metallicity as a function of age for the stellar populations
\citep{2005ApJ...632..872R,2006ApJ...638..585F}. 
The dark matter halos are assumed to follow an NFW density profile 
whereas, the stellar matter is assumed to follow a Kings profile 
embedded within the dark matter halos. 
To simulate Kings profile within dark matter profiles the dark matter 
particles are assigned a stellar mass which is a function of the 
energy of the dark matter particles \footnote{We actually use the massless 
test particles which have 10 times the phase space resolution of the original 
dark matter particle (see \cite{2005ApJ...635..931B} for further details).}. 
The stellar masses depend upon  the adopted star formation time scale 
and the later is chosen by requiring that the simulated satellite
stellar distributions reproduce the structural properties of Local
Group dwarf galaxies.
The three main model
parameters of an accreting satellite are the time since accretion, $t_{\rm
  acc}$, its luminosity, $L_{\rm sat}$, and the circularity of its orbit,
defined as $\epsilon=J/J_{\rm circ}$ ($J$ being the angular momentum of the
orbit and $J_{\rm circ}$ the angular momentum of a circular orbit having same
energy). The distribution of these three parameters describes the accretion
history of a halo. To study the sensitivity of the properties of structures in
the stellar halo to accretion history, additionally a set of six
artificial stellar halo models (referred to as non-$\Lambda$CDM halos)
were generated by \cite{2008ApJ...689..936J}. These have accretion
events that are predominantly
(i) {\sl radial} ($\epsilon<0.2$), (ii) {\sl circular} ($\epsilon>0.7$), (iii)
{\sl old} ($t_{\rm acc}>11 \Gyr$), (iv) {\sl young} ($t_{\rm acc}<8
\Gyr$), (v) {\sl high-luminosity} ($L_{\rm sat}>10^7 \Lsun$), and (vi) {\sl low-luminosity} ($L_{\rm sat}<10^7 \Lsun$).

\subsection{Analytic smooth stellar halo}
For some applications it is also necessary to have a smooth stellar
halo. We do this by assuming an oblate power law distribution with 
ellipticity $\epsilon=0.76$ and power law index of $n_{\rm H}=-2.44$ as 
suggested by R03. 
The values of $\epsilon$ and $n_{\rm H}$ can be altered if desired, e.g., 
recent results by \citet{2008ApJ...673..864J} using SDSS predict 
a flatter halo ($\epsilon=0.64$) and steeper value for the power
law index ($n_{\rm H}=-2.77$).
In accordance with the 
results of  \citet{2010ApJ...716....1B}, the principal axes of the 
velocity ellipsoid are assumed to be aligned with a spherical
coordinate system and the velocity dispersions are assumed to be
 $\sigma_r=141 \kms$ and $\sigma_{\theta}=\sigma_{\phi}=75 \kms$. 
The spheroid is assumed to have no net rotation.

\subsection{Initial mass function and normalizations}
Having specified the density functions, we now provide the 
the initial mass function $\xi$ and the normalization constants like 
$\rho_c$ and the star formation rate $\Psi(\tau)$. 
As in the Besan\c{c}on model, the IMF was assumed to be of a 
power law form, $\xi(m)=m^{-\alpha}$, the values of $\alpha$ are
tabulated in \tab{tb1}. The mass limits of the IMF were defined to be 
in the range  $0.07<m<100$. 
Next, the density normalization constants were chosen to 
reproduce the local mass density of current stars $\rho_0$ 
for the various galactic 
components in the solar neighborhood (Table 2 in R03). 
This resulted in a star formation rate of $\Psi=2.85 \Msun/{\rm yr}$ 
and $\rho_c$ of $1.55\times 10^9$ and $1.31 \times 10^7 \Msun {\rm
  kpc^{-3}}$ for the thick disc and the stellar halo respectively. 
However, for the thin disc in the regime $7<\tau<10 \Gyr$ the SFR 
had to be lowered by 20\% in order to match the star counts in the  
Besan\c{c}on model. This is because in this regime 
the isochrones employed by us differ from those in 
the Besan\c{c}on model. For the bulge  $\rho_c=3.49 \times 10^9 \Msun
{\rm kpc^{-3}}$ was selected which gives a central 
density of $13.76 {\rm \ stars \ pc^{-3}}$ in accordance with the
Besan\c{c}on model.

For our analysis, 
we assume the Sun to be located at 
a radial distance of $8 \kpc$ from the center of the galaxy 
\citep{2003ApJ...597L.121E,1993ARA&A..31..345R} 
Since our adopted radial distance of Sun is different from the Besan\c{c}on 
model ($8.5 \kpc$), we need to recompute the SFR. 
The re-calibrated value of SFR for our adopted
location of the Sun turns out to be $2.37 \Msun/{\rm yr}$.
Also, to match the star counts for the stellar halo at 
large distances (along the north galactic pole) 
with those of the Besan\c{c}on model, 
we had to increase the local mass density of the stellar halo 
by $10\%$.

\section{Tests and results}
In this section we test  the code by comparing its predictions 
with various observational constraints.

\begin{figure}
  \centering \includegraphics[width=0.50\textwidth]{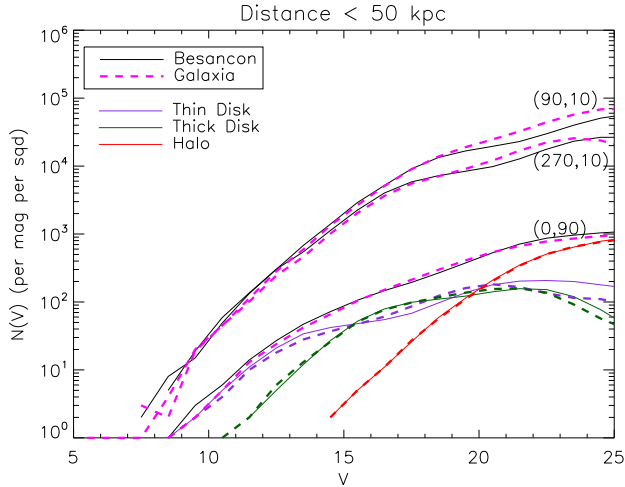}
\caption{Star count predictions of {\sl Galaxia} in various directions compared 
with those of the Besan\c{c}on model. For the direction along the North
pole the contributions for different galactic components are shown separately.
\label{fig:besancon_b90}}
\bigskip
\end{figure}

\subsection{Comparison with Besan\c{c}on} \label{sec:bes_comp}
The star count predictions of the
Besan\c{c}on model have been tested against a variety of observations, and 
since {\sl Galaxia} uses the same disc model, it suffices for most
situations to show that {\sl Galaxia} reproduces the Besan\c{c}on model.  
In order to keep the test simple dust extinction was neglected. 
To test the correspondence of our results with that of 
Besan\c{c}on, we plot in \fig{besancon_b90} the star
counts as a function V band magnitude along three directions; the north
galactic pole, $(l,b)=(90 \degree,10.0 \degree)$ and $(l,b)=(270\degree,10.0\degree)$. 
The last two directions were chosen to illustrate the effect 
of warping which results in a bifurcation in the star counts at
fainter magnitudes.
Overall we find good agreement with the curves from {\sl Galaxia}
corresponding closely to the Besan\c{c}on curves. 

For the direction along the galactic pole, the star counts 
for each of the galactic components are also shown separately. 
It can be seen that the thick disc and the stellar halo are
in perfect agreement but the thin disc 
shows some subtle differences. For the {\sl Galaxia} thin disc, there is 
a slight excess of intermediate magnitude stars and a shortage of
extremely faint magnitude stars. This same effect can be seen in the
other two directions also but is shifted to even fainter magnitudes.
These discrepancies are due to the differences in the isochrones 
employed by us as compared to those used by
R03. Specifically for the thin disc  
in the mass range $0.15<m<0.6$, the absolute magnitude of stars 
predicted by the Padova isochrones were found to be brighter than 
that of the isochrones used by the Besan\c{c}on model.

\begin{figure}
  \centering \includegraphics[width=0.50\textwidth]{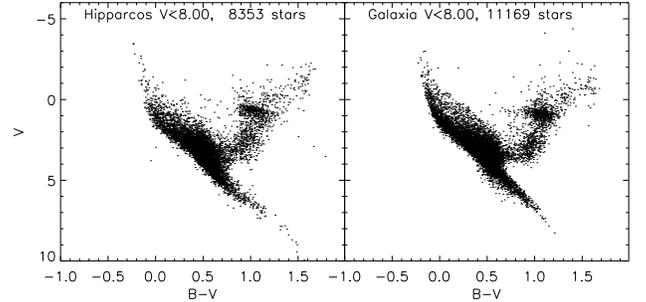}
\caption{Comparison of color-magnitude diagram obtained from the
  Hipparcos catalog (left panel) with that of simulation from
  {\sl Galaxia} (right panel). The plots 
  show stars in the solar neighborhood defined by $V<8$ and 
distance $r<100$ pc. 
\label{fig:hipparcos_cmd}}
\end{figure}
\begin{figure}
  \centering \includegraphics[width=0.50\textwidth]{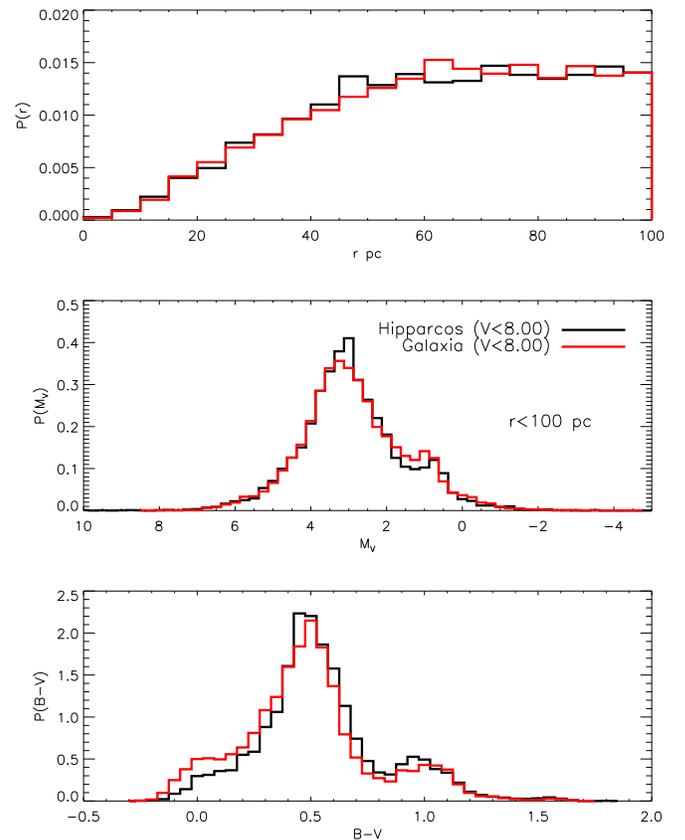}
\caption{Comparison of distributions of radial distance, absolute
  magnitude and color for stars obtained from the
  Hipparcos catalog with that of simulation from
  {\sl Galaxia}. 
\label{fig:hipparcos_rdist}}
\end{figure}

\subsection{Comparison with the Hipparcos data}
The predictions of stellar properties in the solar neighborhood is 
an essential test for any galactic model. For these stars, accurate 
distances can be obtained, which enables the construction of the 
color-absolute magnitude diagram of stars. The Hipparcos mission
\citep{1997yCat.1239....0E} produced an astrometric data base of 117955 stars down to a 
limiting magnitude of $V \sim 12.4$.   For stars with $V<9$, the data 
have a median precision of about 1 mas which is ideal for analyzing
the solar neighborhood. 
The completeness limit of the Hipparcos catalog is estimated to be 
between 7.3-9 mag in V band. Hence we use the following criterion to  
extract a volume complete sample from the Hipparcos catalog \citep[new
reduction]{2007A&A...474..653V} ---
$V<8$ and parallax $\pi>10$ mas. Binary stars were excluded 
from the analysis. 
The Hipparcos magnitude, $H_P$, was converted to $V$ band magnitude 
using the relation $H_P=V+0.408(B-V)-0.13(B-V)^2$ from 
\citet{1992ESASP1136.....T}.
Next, we use {\sl Galaxia} to generate a similar catalog. The parallax errors 
for Hipparcos, which are a function of apparent magnitude, were
simulated as in \citet{2005A&A...436..895G}, where such an analysis 
was earlier performed.
 Although, \citet{2005A&A...436..895G} had not taken extinction into
account we here do take it into account. 
For this study we adopt the extinction model presented
in \citep{2009MNRAS.397.1286A} such that,
\begin{equation} 
E(B-V)={\rm Max}(0,0.47(d-0.07 \kpc)), 
\end{equation}
which takes into account the fact that the extinction 
within 70 pc from the Sun is negligible.
The reddening was converted to extinction assuming an $R_{V}$ of 3.1. 
Note, the extinction is quite small, and hence it is not expected 
to have any significant impact on the results.

\fig{hipparcos_cmd} shows the color magnitude distribution of stars 
within  100 pc from the Sun using the Hipparcos and the simulated data
sets. The plots are very similar. In \fig{hipparcos_rdist} the probability 
distributions of absolute magnitude, color and distance  are also
shown. All of the distributions show good agreement with the Hipparcos  
data. Some minor differences do exist, in particular:  
(i) In the radial distribution of stars there is a slight excess of 
stars at $r=45$ pc and this  
is due to the Hyades cluster as pointed out by
\citet{2005A&A...436..895G}. However, unlike
\citet{2005A&A...436..895G} our overall radial distribution is in good 
agreement with Hipparcos and does not show a deficit at low $r$ and 
excess at high $r$ as reported by them. 
(ii) The absolute magnitude distribution of {\sl Galaxia} 
shows a mild excess at $M_V\sim 1$ and a slight deficit at $M_V\sim
3$. This excess was also observed by \citet{2005A&A...436..895G} 
and is of comparable magnitude. Again, unlike 
\citet{2005A&A...436..895G} we do not see any deficit in the 
faint tail ($M_v>4$).  
(iii) In the color distribution, the red clump peak at $B-V \sim 1$ 
is shifted slightly to the red side for {\sl Galaxia} and the blue 
tail shows a slight excess of stars. As remarked by
\citet{2005A&A...436..895G} these discrepancies could be due to 
imperfections in the stellar evolution models or due to imperfect 
simulation of the parallax errors and should be investigated in
future.

Other than the minor differences discussed above, there is one major 
difference between the two catalogs. The number of stars in the simulated 
catalog is about 33\% higher. This is due to the fact that we have 
excluded the stellar multiplicity (e.g. unresolved binary systems)  from the Hipparcos
catalog. For the selection limits used by us the Hipparcos 
catalog contains about 1549 sources that are flagged as binary. 
Multiplying this number by two and adding it to the number of 
single sources (8353) we get a total of about 11451 stars, 
which is in good agreement with the number of stars in the 
simulated catalog (11169).

\begin{figure}
  \centering \includegraphics[width=0.50\textwidth]{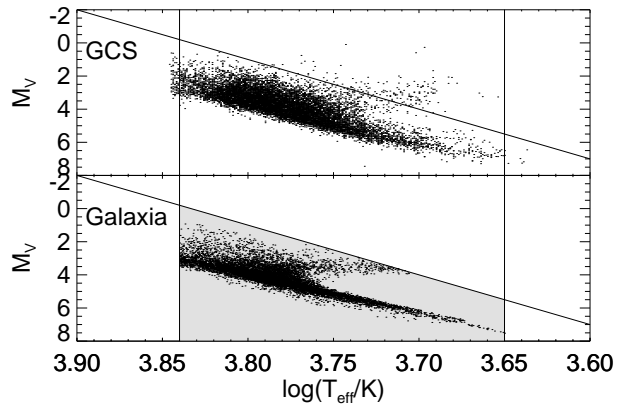}
\caption{ Distribution of GCS stars in the $(T_{\rm eff},M_V)$ plane 
and the choice of selection functions. The top panel shows 
the GCS stars lying within $r<0.12\kpc$ and the lower panel 
shows the stars sampled with {\sl Galaxia}. The shaded region 
in the lower panel shows the selection region used to mimic 
the GCS stars. The line represents the equation 
$M_V=115-30\log(T_{\rm  eff})$.
\label{fig:gcs2}}
\end{figure}

\subsection{Comparison with solar neighborhood kinematics from the Geneva-Copenhagen survey}
The best kinematic data to date are largely restricted to the solar 
neighborhood. 
We now compare these observations with kinematic simulations 
from {\sl Galaxia}. 
We use the data from the Geneva-Copenhagen Survey, GCS,  
\citep{2004A&A...418..989N,2009A&A...501..941H}, which is a selection
of 16682 F and G type main sequence stars, out of which  
velocities and temperatures are available for 13382 stars. 
The GCS catalog is complete only 
till $r \sim 40 {\rm pc}$ in volume and  $V \sim 8$ in magnitude and 
within these limits there are only about 1000 stars. Hence, to 
increase our sample size we select stars within  120 pc 
from the Sun and with $V<9.6$, 10893 stars were found to satisfy 
this criteria. A drawback of this is that beyond the completeness 
limit of the survey, it is not possible to deduce the exact selection 
functions and these are needed to mimic the GCS catalog with {\sl Galaxia}.
However, the incompleteness was found to have 
little effect on the kinematic properties of the sample.   
To check this we compared the velocity distributions of the complete 
and incomplete samples as described above and found them to be 
nearly identical. Hence, an approximate selection function 
that can reproduce the main properties of the GCS sample, namely, 
the selection of F and G type dwarfs with minimal contamination 
from red giants,  should be sufficient for our purpose. 
To do this, we plot in \fig{gcs2} the distribution 
of our selected GCS stars, in the $(T_{\rm eff},M_V)$ plane. 
The shaded 
region in the plot shows the desired selection function in the $(T_{\rm
  eff},M_V)$ plane, which can be described by the following equations 
$3.65<\log(T_{\rm eff})<3.84$ and $M_V<115.0-30\log(T_{\rm eff})$.
We use this to reproduce the GCS catalog with
{\sl Galaxia} and this is shown in the lower panel. 
The stars generated with {\sl Galaxia}  were also restricted to  
$V<9.6$ in magnitude and  $r<120 {\rm pc}$ in distance.

\begin{figure}
\centering \includegraphics[width=0.50\textwidth]{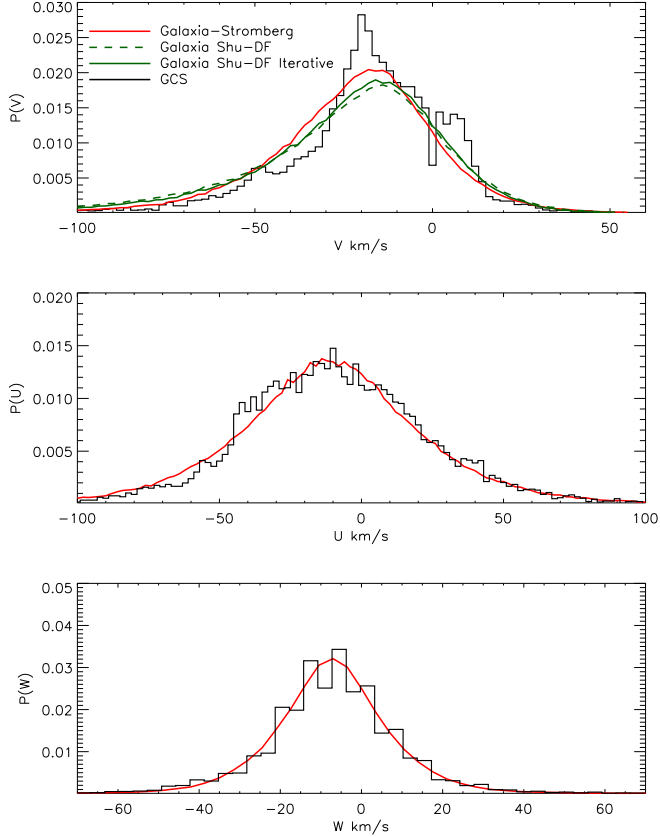}
\caption{ Predicted velocity distributions near the Sun. Shown are the
U, V and W component of velocities. The dotted lines show the 
distributions of F and G type stars in the GCS survey of the 
solar neighborhood  \citep{2004A&A...418..989N} ($r<0.12\kpc$).
\label{fig:f1}}
\end{figure}

Next, in \fig{f1}, we plot the predicted distributions of 
the U, V and W component of velocities and compare them to the GCS
survey. It can be seen that the model (the red curves) 
provides a reasonable fit to the distributions of all the 
three components of velocities. In the plots shown in the panels,
the Sun's motion with respect to the local standard of rest (LSR)
was assumed to be $U_{\rm \Sun}=11.1$ $V_{\rm \Sun}=12.24$ and 
$W_{\rm \Sun}=7.25$ as given by \citet{2010MNRAS.403.1829S}. 

In {\sl Galaxia} we model the distribution of the $v_{\phi}$ component
of velocities 
using the asymmetric drift relation given by \equ{stromberg}. 
As mentioned earlier,  an alternative way to model the $v_{\phi}$
velocities is to use the Shu distribution function  
which provides a more accurate treatment of the radial motion in the
disc (see Appendix \ref{sec:rotational_kin} for further details).
In \fig{f1} this is shown as the dashed curve. 
The Shu distribution function (DF)
is found to overestimate the negative tail slightly and its peak is 
also slightly to the right as compared to the one predicted by \equ{stromberg}. 
Using the iterative scheme of \citet{1999AJ....118.1201D} one can
further  improve the accuracy of the solution (dark green solid line
in \fig{f1}), but even this has only a marginal effect in resolving
the overestimation of the negative tail. 
On the other hand, 
lowering $\sigma_{r0}$ for both the thin and thick disc was found to 
reduce this discrepancy  but this also makes the 
$U$ distribution sharper than that observed for GCS stars.  
It should be noted that in the regime $V<-60\kms$ where 
the tail is overestimated, Poisson errors are also high, 
owing to low number of stars, hence we refrain from trying to 
alter the model to fit the data at this stage.  
Note, the fact that the Stromberg relation provides a better 
fit to the negative tail than the Shu DF, is only because it 
underestimates the negative 
tail as compared to the distribution expected when radial motions 
are properly taken into account. 

Finally, in a manner similar to \citet{2010MNRAS.403.1829S}. 
our model distributions can also be used to calculate 
the solar motion with respect to the LSR. 
To accomplish this we applied a $\chi^2$ minimization scheme 
and fitted our model $U,V,W$ distributions to the distributions 
of the GCS stars (a uniform weighting scheme was used). 
We found $(U_{\rm \Sun},V_{\rm
  \Sun},W_{\rm \Sun})=(11.25\pm 0.3, 10.93\pm 1.2,7.86 \pm 0.3) \kms$  
($1 \sigma$ range), which are in good agreement with values of
\citet{2010MNRAS.403.1829S}. 
However, our predicted  value of $V_{\rm
  \Sun}$ is closer to value reported by 
\citet{2010MNRAS.401.2318B} ($11 \kms$) than that of 
\citet{2010MNRAS.403.1829S}. 
Note, this result is with the $V$ component of velocity being
modeled using the Shu distribution function, if instead the 
relation given by \equ{stromberg} is used, the best fit value of 
$V_{\rm \Sun}$ is found to be even lower ($\sim 10 \kms$). 
However, this discrepancy is not significant considering the fact that
a random uncertainty of about $1 \kms$ is associated with it. 
Moreover, the substructures being quite dominant in the distribution 
a systematic uncertainty of as high as 
$2 \kms$ is also expected \citep{2010MNRAS.403.1829S}.

 \begin{figure}
  \centering \includegraphics[width=0.50\textwidth]{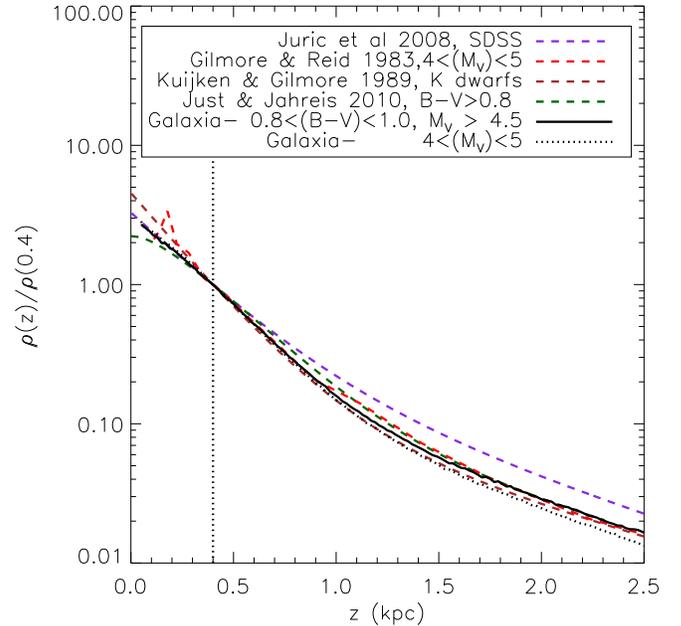}
\caption{ Vertical distribution of star counts using {\sl Galaxia} and 
its comparison with published results. 
\label{fig:f2}}
\end{figure}

\subsection{Vertical distribution of stars in the solar neighborhood}
Predictions for vertical distribution of stars is another important
test for a model. Traditionally the vertical distribution is modeled 
as a sum of two exponentials representing the 
thin and the thick disc. The best fit parameters 
as provided by \citet{2008ApJ...673..864J} for the M dwarfs in SDSS,  
after correction for biases, e.g., stellar multiplicity, 
are a thin disc with a scale length of 300 pc, a thick disc 
with a scale length of 900 pc, and a local thin to thick disc 
normalization of $f=13\%$. Using K dwarfs towards the south galactic
pole (SGP), \citet{1989MNRAS.239..605K} report a value of $f=4.27\%$ and scale
lengths of 249 pc and 1000 pc for the thin and thick discs, respectively. 
Also using the data towards the SGP, 
\citet{1983MNRAS.202.1025G} provide the vertical distribution of stars
for different absolute magnitude ranges --- 
we use results reported for magnitude range 
$4<M_V<5$ since stars at higher absolute magnitudes could be contaminated 
by giants. \fig{f2} shows these published results alongside 
our results using {\sl Galaxia}.
Also is shown the results from 
the model of \citet{2010MNRAS.402..461J}. 
Note, we normalize all the profiles to the density at $z=0.4 \kpc$. 
This is because density estimates close to $z=0$ 
cannot be reliably obtained and one has to use interpolation 
to get the local normalizations, which makes the estimates strongly 
dependent upon the assumed shape of the profile 
\footnote{Interestingly, a compilation from literature of local normalization 
versus scale height by \cite{2009IAUS..254P...5A} seems to suggest that, 
in spite of a large spread in the values of these quantities,
the estimated total mass between different surveys is roughly the same.}.
Although using the SDSS selection functions would have been 
ideal for comparing with the predictions of {\sl Galaxia}, the theoretical 
isochrones are not very accurate in the M dwarf regime at the present time. 
Hence, we use a color range of $0.8<B-V<1.0$ and $M_V>4.5$ to identify the 
main sequence dwarfs and report the results for them. 
Our results correspond closest  
to the plot of \citet{1983MNRAS.202.1025G}. Closer to the 
plane ($z<0.5\kpc$) they are also in good agreement
with that of \citet{2008ApJ...673..864J}, but at larger distances 
they differ. 
We also show the results with stars selected according to their 
absolute magnitude ($4.0<M_V<5.0$). This is quite close to the 
color limited profile but differs slightly in the regime $z>1$.
This shows the effect of selection bias on the profile.
Note, our assumed location of the Sun above the galactic
plane (15 pc) is slightly less than that reported by 
\citet{2008ApJ...673..864J} (25 pc), but this has negligible impact 
on the analysis presented here.

\begin{figure}
  \centering \includegraphics[width=0.50\textwidth]{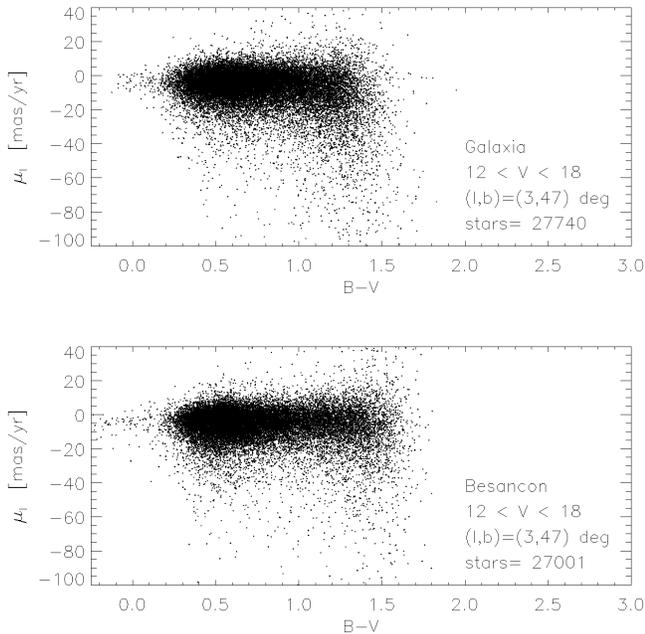}
\caption{Plots of color $B-V$ vs proper motion $\mu_l$ towards
  $(l,b)=(3,47)^{\degree}$, for $12<V<18$. Upper panel shows results
  obtained with  {\sl Galaxia}  while the lower panel shows results 
  from the Besan\c{c}on model.
\label{fig:proper_motion}}
\end{figure}

\subsection{Color vs proper motion comparison}
In R03, distributions of color vs proper
motion from Besan\c{c}on model were shown and found to be in good 
agreement with observations of \citet{1996A&A...311..456O}. 
Here we repeat the same analysis and compare our results 
with those of the Besan\c{c}on model (\fig{proper_motion}). Stars were 
identified in a 15.5 square degree area in the direction of  
$(l,b)=(3,47)^{\degree}$ and with magnitudes in the range $12<V<18$. 
In upper panel of \fig{proper_motion} the bluer region is dominated by the 
thick disc and the halo, while the redder region is dominated by the 
thin disc. It can be seen that our distribution is in good agreement with 
that of the Besan\c{c}on model. However, subtle differences can 
be identified in the region $1<B-V<1.5$. At around $B-V \sim 1.0$,
the dispersion of  $\mu_l$ is slightly less  for Besan\c{c}on. Also, the  
Besan\c{c}on stars extend further redwards. 
These differences owe their origin 
to the differences between the isochrones used by us and the 
Besan\c{c}on model (see \sec{bes_comp}).

\subsection{Computational performance}
We have implemented the scheme in the form of a serial code written in
C++. It uses about 500 MB RAM and the amount of memory used 
is independent of the size of the catalog being simulated. 
However, the CPU time depends sensitively upon the parameters 
of the survey being simulated. 
In what follows, we report CPU 
times by running the code on a 2.44 GHz single Intel processor.

Recall that in our scheme the Galaxy is divided into roughly 
equal mass cubical boxes (nodes), and stars are generated from only 
those nodes that intersect with the survey volume. 
This causes the run time to vary nearly linearly with the mass of the 
galaxy being sampled by the survey.
Now, for a given survey there are two main parts 
that consume most of the CPU time: a) time to pre-process a node, and 
b) time to generate stars in a node. 
Pre-processing all of the nodes 
takes about 250 seconds. For surveys sampling a smaller mass fraction, this is 
also proportionately lower. Generating stars is where the code 
spends most time and to measure this we define the 
speed of the code as the number of stars generated per 
second.  The code can reach a maximum speed of 0.2 million stars per 
second when generating all stars in the galaxy.
For magnitude limited surveys, the speed is slightly lower, e.g., 
a $V<20$ survey runs at 0.16 million stars per second while a 
$V<15$ survey runs at 0.05 million stars per second.
The optimization of the code to generate only stars that can 
enter the catalog is imperfect, especially if a galactic component 
has a large spread in metallicity. This is something we hope 
to improve in future. 

With the speeds given above, an all-sky GAIA-style
survey with $V<20$, consisting of 
4 billion stars can be generated in about 6.5 hours. 
In comparison, the scheme of \citet{Brown2005}  
required about 2 weeks to generate a 
GAIA like sample of $3.5 \times 10^8$ stars while 
running on a cluster of processors. 
On the other hand, a $V<20$ survey confined to 10,000 sq degree 
towards the north galactic pole, consisting of a sample of $3 \times 10^7$
stars can be generated in 220 seconds. 

Note, running a single instance of the code is 
sufficient for generating 
catalogs of sizes up to $10^8$ stars in less than half an hour, 
but for larger catalogs one could easily parallelize the code.
Parallelization simply involves running multiple 
instances of the code, each with a different random seed and 
generating a fixed fraction of the full galaxy. 

\begin{figure}
  \centering \includegraphics[width=0.50\textwidth]{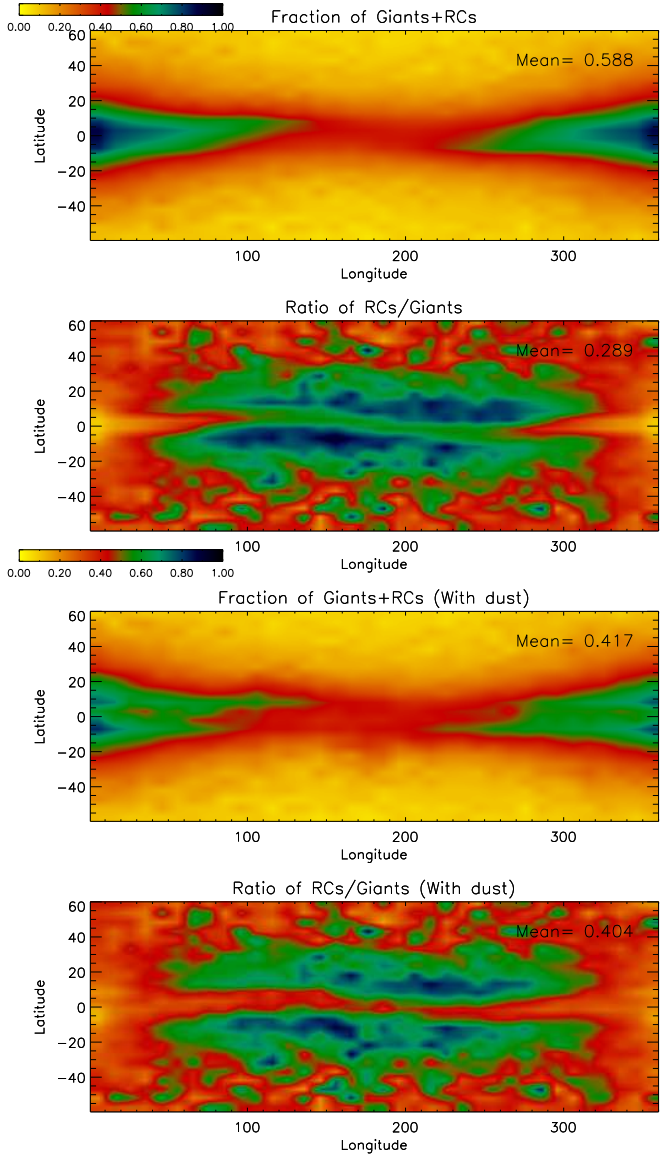}
\caption{ Angular distribution of giants and red clumps. Shown are the 
fraction of giants and red clumps, and the ratio of red clump to 
giants in a survey with magnitudes in the range $9<I<13$. 
The upper panels are without
dust extinction while the lower panels are with dust extinction.
In the figures, the small scale structures lying away from the 
galactic mid plane are due to Poisson noise resulting from low number of stars.
\label{fig:rave_giant}}
\end{figure}
\begin{figure}
  \centering \includegraphics[width=0.50\textwidth]{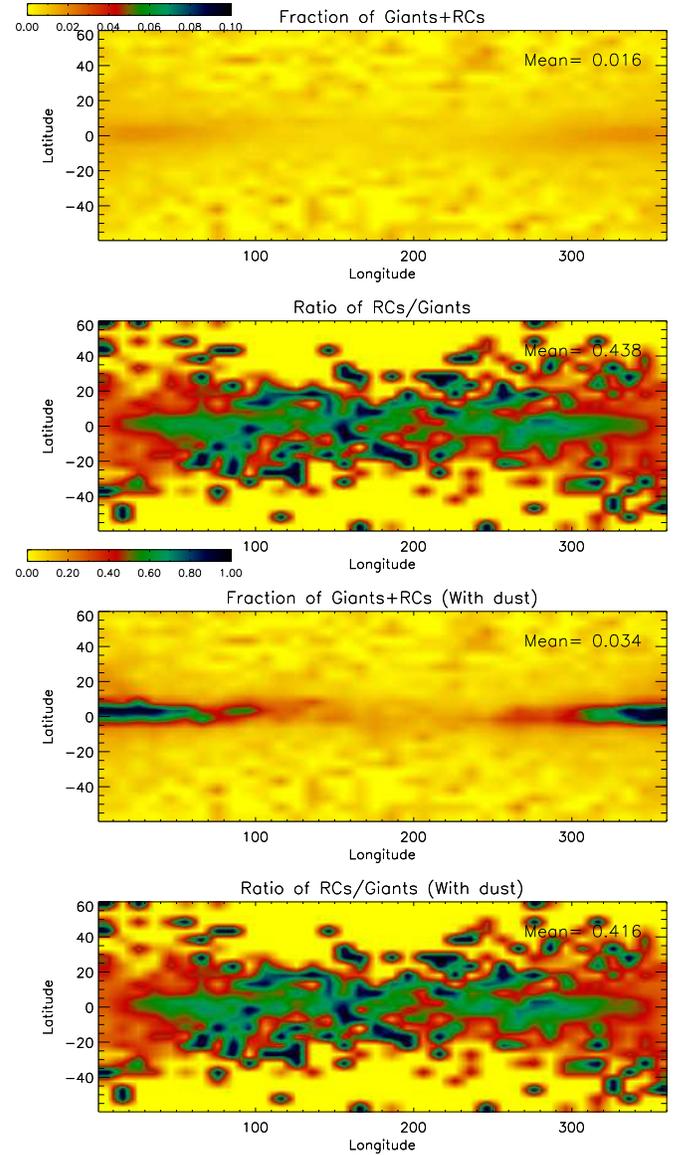}
\caption{ Angular distribution of giants and red clumps. Shown are the 
fraction of giants and red clumps, and the ratio of red clump to 
giants in a survey with SDSS $r$ band magnitude in the range $4<r<22$ . 
The upper panels are without
dust extinction while the lower panels are with dust extinction.
The color bar range for 1st and 3rd panels is 0 to 0.1 and for  
2nd and 4th panels is 0 to 1. 
In the figures, the small scale structures lying away from the 
galactic mid plane are due to Poisson noise resulting from low number of stars.
\label{fig:sdss_giant}}
\end{figure}

\section{Applications}
\subsection{All-sky distributions of giants and red clumps}
As an application of the code, we investigate the distribution 
of giants and red clump stars over the sky. 
We use the following criteria to identify the stars that are 
giants and red clumps;  $T_{\rm  eff}<$5000K and $\log(g) < 3.3$. 
A subset of these stars having  4500K$<T_{\rm  eff}<$5000K and 
$1.625 <\log(L/\Lsun)< 1.825$ were identified as red clumps. 
\fig{rave_giant} shows  all-sky maps for the fraction of stars that
are giants and red clumps, and the ratio of red clump to giants, for a RAVE-style survey with 
magnitudes in the range $9<I<13$. The lower panels show the same 
but with dust extinction included. 
Except for the effect of the warp 
the plots being roughly symmetrical about $l=180\degree$,
we concentrate on the first half only, i.e., $0\degree < l< 180\degree$. 
It can be seen in \fig{rave_giant} that the fraction of stars that are
giants and red clumps decreases as a function of 
$l$ and $|b|$ as one
moves away from the galactic center and the galactic mid plane
respectively. This is mainly due to the fact that 
giants being more luminous can be seen farthest 
in a magnitude limited sample, and for a given solid angle, the volume 
sampled increases as cube of the distance. Hence, in directions where 
the disc extends the farthest, we should expect to see a
higher fraction of giants.  Red clumps on the other hand 
are less luminous than the giants which means that the ratio 
of red clumps to giants will be lower in regions where the fraction 
of giants is higher. Additionally, the red clumps are found in metal
rich populations only. Now, due to the existence of a vertical
metallicity gradient in the disc, as one moves away from the galactic
plane the fraction of metal rich population being surveyed decreases, 
this makes the ratio of RCs/giants to fall off as one moves away 
from the galactic mid-plane. 
Including dust extinction 
in general leads to faint stars getting excluded in a volume limited survey. 
Since a RAVE-style survey has a significant fraction of giants 
lying at larger distances, where they also appear faint, the 
effect of extinction is to lower the overall fraction of giants 
(from 0.6 to 0.4). 
As expected the overall ratio of
red clump stars to giants increases slightly (from 0.3 to 0.4) 
on including extinction. 

Next, we repeat our analysis for an SDSS-style survey,
with $r$ band magnitude in the range $4<r<22$ (\fig{sdss_giant}). 
The mean fraction of stars that are giants and red clumps 
is found to be 0.016, which increases   
to 0.034 when extinction is included. The mean ratio of red clumps 
to giants
is 0.44, which with extinction included decreases only slightly to 0.42. 
Here the effect of extinction on the fraction of giants is opposite to 
what we saw earlier for the RAVE simulation. This is due to the fact  
that in an SDSS type survey the faint end of the magnitude distribution 
is dominated by main sequence dwarfs, which get excluded by the effect
of extinction. 
On the other hand, given the extremely faint magnitude limit of the
survey,  the giants and red clumps in the disc are 
luminous enough to be visible even after including the effects of 
extinction.

\subsection{Identification of substructures in the stellar halo in
  $E-L_z$ space with GAIA} \label{sec:gaia_anal}
\begin{figure}
  \centering \includegraphics[width=0.50\textwidth]{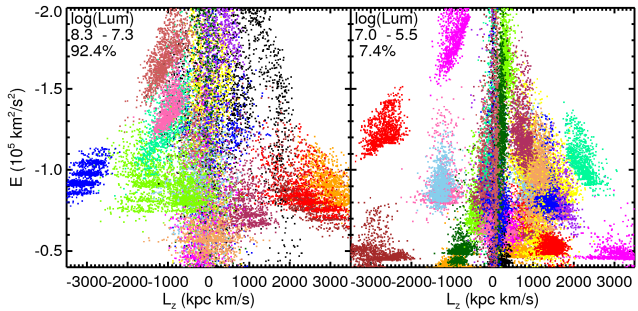}
\caption{Distribution in $E-L_z$ space  of particles lying within 150
  kpc from the galactic center in one of the BJ05
  simulations. 17 unique colors were used with each color representing
 a different satellite. Left
  panel shows the 20  most luminous satellite systems while the right 
 panel shows the next 40 most luminous systems. For each panel the range in 
 luminosity in units of $\log(L/L_{\odot})$ and its percentage
 contribution to the total halo is also labelled. 
 Each satellite system is sampled with about 1000 
 particles and the sampling probability of  a particle with in a given
 satellite is proportional to its luminosity. The satellites were
 sorted in decreasing order of their luminosity before plotting, so as
 to  make the lower luminosity satellite systems, which are expected
 to cover a smaller area in the map, lie on top of higher luminosity  systems.
\label{fig:gaia_4}}
\end{figure}
As an application of our code, we assess the ability of the
astrometric mission GAIA to identify structures in
the stellar halo. 
\citet{2000MNRAS.319..657H} first performed such an analysis and
concluded that given the accuracy of GAIA, it would be relatively easy 
to identify the structures in integrals of motion space, e.g., 
energy $E$ vs the $z$ component of angular momentum $L_z$. Subsequently,
\cite{Brown2005} showed that in the $E-L_{z}$ space favored
by \citet{2000MNRAS.319..657H}, the 
structures are nearly washed out if the background population of stars 
are taken into account.  While \citet{2000MNRAS.319..657H}
assumed the halo to be formed out of in-falling satellites exclusively, 
\cite{Brown2005} superimposed a few disrupted satellites onto an
otherwise smooth stellar halo. Hence neither of the simulations were 
done in a proper cosmological context. Recently \cite{2010MNRAS.tmp.1196G}
(GHBL hereafter) have done higher resolution simulations similar to 
\citet{2000MNRAS.319..657H} along with some enhancements, in
particular, a time
dependent analytic potential, and satellite luminosities drawn from the
present-day distribution of Milky Way satellites
\citep{2008ApJ...686..279K}. They conclude that significant amount 
of structure can be seen in the $(E,L_z,L)$ space. 
We repeat a similar analysis here but using the N-body models of
\citet{2005ApJ...635..931B} that simulate the halo in a
cosmological context and hence have realistic time-dependence of rate
of in-fall and luminosities of accreted satellites.

First we investigate as to how clustered are the particles 
of a tidally disrupted satellite in $E-L_z$ space and compare them 
with the results of GHBL. Specifically we reproduce Figure 4 of GHBL  
showing the distribution in $E-L_z$ space  of particles 
in the simulations with different colors representing 
different satellites. Results from one of the BJ05 simulations 
used by us are shown in \fig{gaia_4}. While GHBL show all
43  satellites simulated by them, we show 20 most luminous satellites 
in left panel and next 40 luminous in right panel. The 20 most
luminous systems which constitute about 92\% of the stellar 
mass are significantly less clustered than the other low luminosity 
systems. The Fig-4 in GHBL in fact resembles the right panel, which means that 
energy and angular momentum is conserved less in BJ05 than in GHBL. 
Also a lot particles can be seen at low  
$L_z$ and high $E$ in BJ05 as compared to GHBL suggesting that
there are more high energy radial orbits in BJ05 simulations. 

Given the differences in the results of GHBL and BJ05, it is imperative to
look at the differences between the methodology of the two simulations, 
so as to isolate the cause. 
\begin{itemize}
\item Dynamical friction is ignored in GHBL
whereas BJ05 use a modified version of the Chandrashekhar 
dynamical friction formula. Since dynamical friction is strongest 
for massive systems this will make luminous systems appear fuzzier,
an effect seen in BJ05 simulations. This could be one of the reasons 
for the BJ05 results being fuzzier in general than GHBL.   
\item In GHBL the stellar particles of a satellite are not 
explicitly embedded in dark matter potentials.
In general one expects embedded satellites to take longer to disrupt and 
moreover they disrupt in parts with each pericentric passage redistributing
the energy and hence leading to more non conservation of energy.  
\item In BJ05 the satellites are
evolved from the time since they were accreted to the host halo whereas 
in GHBL all satellites are evolved for 10 Gyrs. As a consequence in BJ05 
one expects older systems to be fuzzier and young systems to be clustered. 
\item In GHBL  the orbit initial conditions  
are drawn from a distribution function  corresponding to the 
density profile of the stellar halo at z=0  whereas in BJ05 they 
are motivated from cosmological simulations. The overall effect of
this on the results is not clear. 
\end{itemize}

\begin{figure}
  \centering \includegraphics[width=0.50\textwidth]{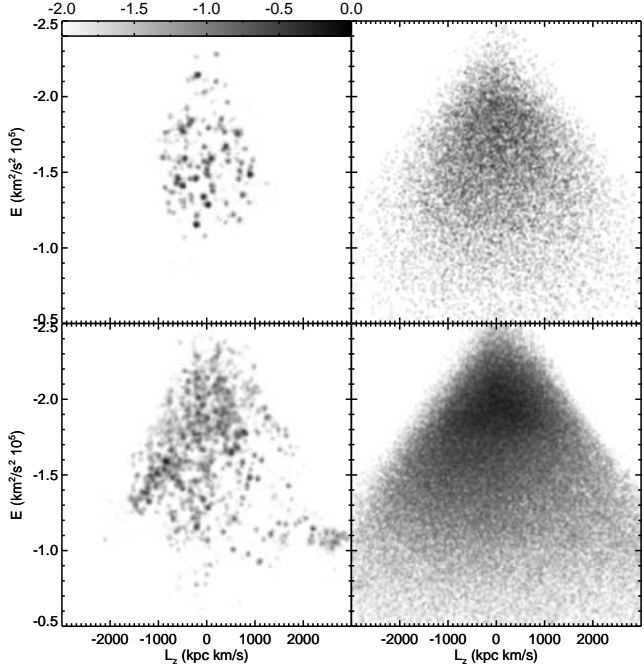}
\caption{Number density map in $E-L_z$ space  of stars lying within
  $4 \kpc$ from the sun in one of the BJ05 simulations. 
Top panels: stars from a single satellite. 
Bottom panels: all stars.
In the left panels, stars have the same phase space coordinate as 
that of their parent N-body particles. This means multiple stars can have 
same phase space coordinates. Hence, for the aid of visibility, a dispersion 
of $30 \kpc\kms$ along $L_z$ and 1000 $\kmstwo$ along $E$ was added to the
position of stars in the left panels. 
The color table  represents $\log(\rho_{\rm max}/\rho)$,  $\rho$ being
the number density.  
\label{fig:gaia_5}}
\end{figure}

Next, we investigate the prospect of detecting the structures in 
the stellar halo with GAIA. 
The parallax and photometric errors of GAIA increase steeply beyond 
$V \sim 15$, hence GAIA will be most accurate for nearby and/or bright
stars. Hence similar to  GHBL we generate a solar neighborhood sample 
of stellar halo stars, using BJ05 simulations,
 with the following constraints-- $V<16$, 
$M_V<4.5$ and $r<4 \kpc$. This resulted in a sample of $1.3\times 10^5$ stars, 
which is similar but slightly larger than that of GHBL ($0.8 \times 10^5$). 
In \fig{gaia_5} we show a density map of stars in the $E-L_z$  space. 
In the left panel the stars have same phase space coordinate as that
of their parent N-body particle, whereas in the right panel the
spawned stars are distributed in phase space according to the scheme 
mentioned in \sec{nbody_samp}. It can be seen that dispersing
the  spawned stars in phase space has a disastrous effect of
smoothing out the distribution of stars. 
In the top panels, we show the distribution of stars 
a single satellite system. It can be seen 
that even in the original N-body coordinates (top left panel) 
the stars are not strongly clustered. Spawning 
stars from it makes the situation even more worse (top right panel). 
The problem stems from the limited numerical resolution of BJ05 
simulations, specially a few massive ones which dominate most of the 
mass in the stellar halo.  
Within 4 kpc from the Sun, typically a satellite 
contributes about 200 particles, and the range of scales
covered in velocity space is of the order of $\pm200 \kms$. 
Now, considering the fact that we smooth over 64 particles in phase 
space, it is easy to see that large amount of scatter is 
introduced when spawning multiple stars from a single particle. 
Additionally, $L_z$ being a product of position and velocity coordinates,  
even a small amount of dispersion in position and velocity 
coordinates can cause a large spread in the value of $L_z$.
Incidentally, GHBL were able to identify significant substructure 
in their analysis. 
However, due to the above mentioned limitations we 
cannot do a faithful  comparison with the results of GHBL.

\begin{figure}
  \centering \includegraphics[width=0.5\textwidth]{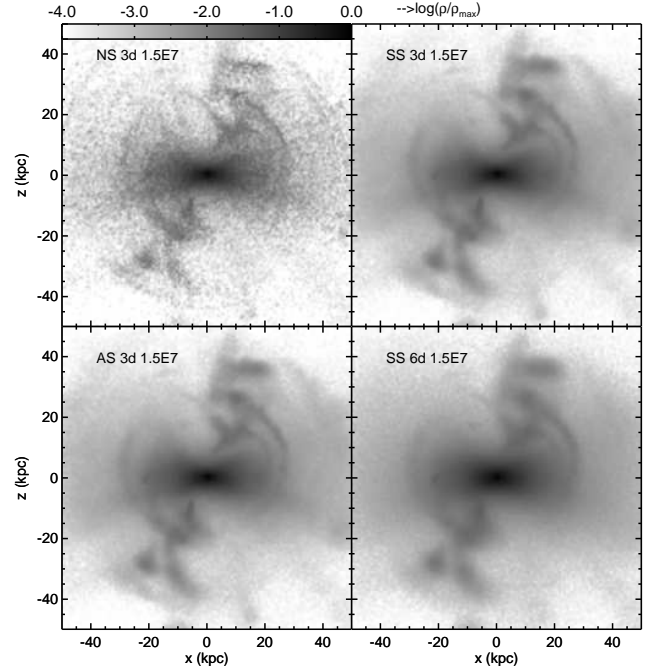}
\caption{Number density map of stars in $x-z$ space 
of a simulated halo sampled by {\sl Galaxia} at a resolution of $1.5
\times 10^7$ with different sampling schemes (NS,AS,SS). 
The bottom right panel is sampled in 6d while the rest of the 
panels  are sampled in 3d.
\label{fig:xz_density}}
\end{figure}

\begin{figure*}
  \centering \includegraphics[width=0.99\textwidth]{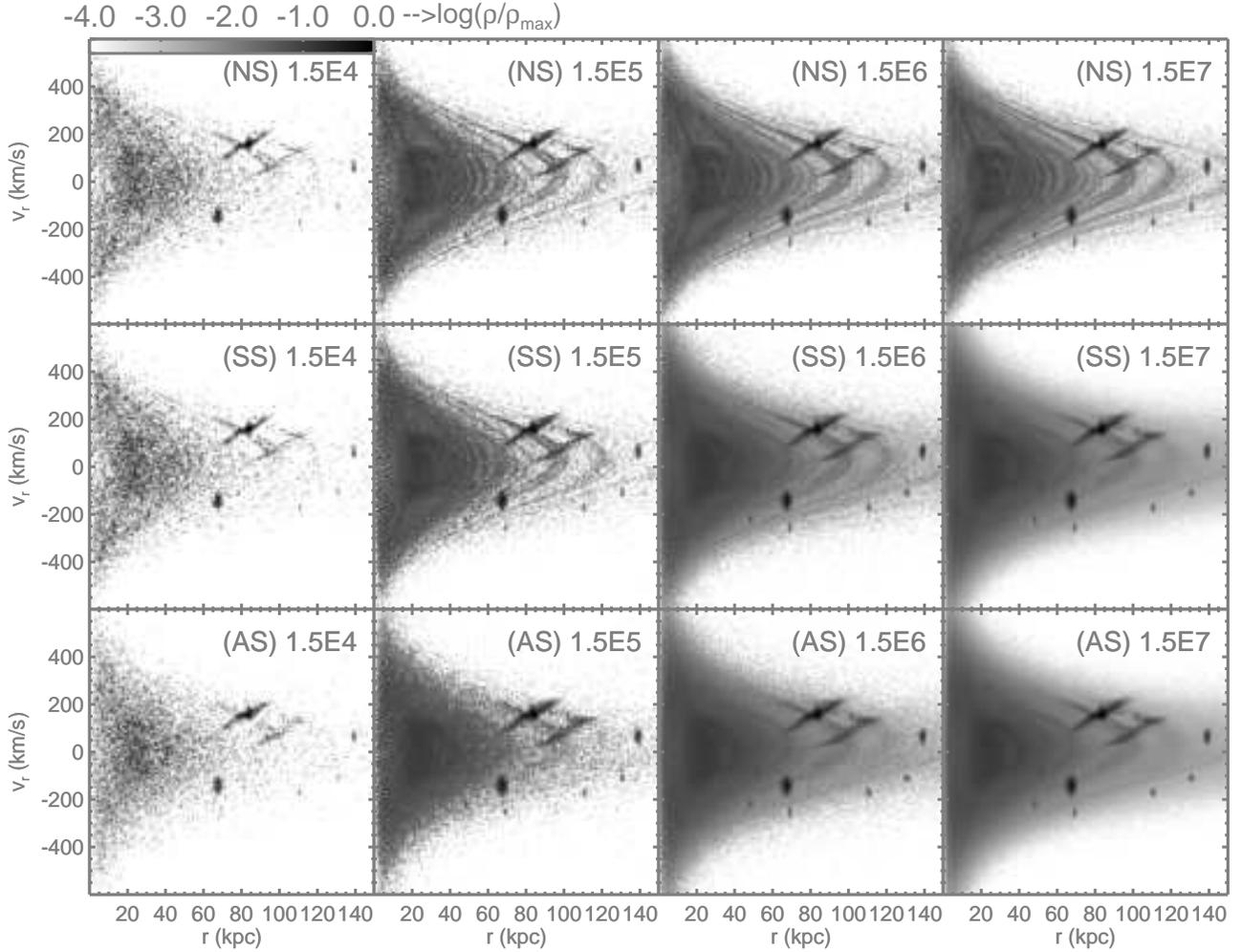}
\caption{Number density map of stars in $r-V_r$  space 
(radial distance and radial velocity with respect to galactic center)
of a simulated halo sampled by {\sl Galaxia} at different resolutions ($1.5
\times 10^4$ to $1.5 \times 10^7$) and
different scattering schemes (NS,AS,SS). Label NS stands for no
scattering scheme, where
the stars are simply assigned the co-ordinates of its parent N-body
particle. Label SS stands for the improved selective scattering
scheme, where the stars are scattered only if its parent N-body particle 
generates more than one star. Finally, label AS stands for the scheme 
where all the spawned stars are scattered.
\label{fig:radial_velocity}}
\end{figure*}

\subsection{Substructures in the stellar halo in $(x,y,z)$ and
  $(r,v_r)$ space} \label{sec:rvr_anal}
According to our discussion in \sec{gaia_anal} it is clear 
that when multiple stars are spawned from a single N-body particle 
significant smoothing of structures occurs. Hence, it is imperative 
to check the regime in which the results are reliable.

In \cite{2010arXiv1012.3515S,2010ApJ...722..750S}, an application of
the code was shown 
for identification of structures in 3d $(x,y,z)$ space making use of the 
photometric information of stars. Note, when kinematics 
are not needed the sampling resolution can be improved by calculating the 
smoothing lengths in 3d position space only. Moreover, a smaller number 
of smoothing neighbors, typically 32, can be used. 
This improves the spatial resolution by a factor of 3 on average when    
compared to smoothing lengths in 6d space with 64 neighbors. 
To check this in \fig{xz_density} we plot the surface density maps 
of stars of a simulated halo as sampled by {\sl Galaxia}. 
A color limit of $0.1<g-r<0.3$ (SDSS $g$
and $r$ band) and magnitude limit of $r<27.5$ was used to sample 
the main sequence stars, so as to fairly sample all the relevant stellar
populations in the halo. A sub-sampling factor of $f_{\rm sub-sample}=0.25$ 
was used which resulted in about $1.5 \times 10^7$ stars. A slice 
with $|y|<10 \kpc$ was selected to make the plots. The 
top left panel 
shows the map as produced when the stars have the same phase space 
coordinates as that of the original N-body particle (no scatter scheme NS). 
The top right panel shows the results with the default  
sampling scheme of {\sl Galaxia} where the stars are scattered only when 
an N-body particle spawns more than one stars 
(selective scatter scheme SS). The bottom left 
panel show the maps when all the spawned stars are scattered
(all scatter scheme AS). The bottom right panel shows the map when 
the stars are sampled in 6d phase space. At such a high sampling 
resolution nearly all N-body particles spawn more than one star, 
hence the SS-3d and AS-3d results are identical. Comparing SS-3d with NS panel 
it can be seen that all the visible structures in NS are 
adequately reproduced by {\sl Galaxia}. However, extremely low contrast 
structures in the outskirts in the form of shells are smoothed out. 
The SS-6d scheme also reproduces the prominent structures correctly, 
but in general its results are slightly more smoothed when compared to 
SS-3d, which is expected as working in higher dimensions erodes the 
spatial resolution. 

Next, we look at the prospect of analyzing the data when radial 
velocity information is also available.
Recently two point correlation function in $(x,y,z,v_r)$ 4d space 
has been used to quantify the amount of structure in the stellar halo
\citep{2010arXiv1011.1925X,2010arXiv1011.1926C,2009ApJ...698..567S}.   
Typically when comparing theory with observations the age and 
metallicity distribution of all the N-body particles is assumed to be 
same. In BJ05 simulations each progenitor has its own stellar 
population and with {\sl Galaxia} it is now possible to draw stars using this 
information. In \cite{2010arXiv1012.3515S} it was shown that depending upon the 
color and magnitude cuts employed the sampling probability is 
different for different accretion events.

In \fig{radial_velocity} we show the number density map of stars in the 
$r-V_r$ space (radial distance and radial velocity with respect to galactic 
center).
The top, middle and bottom panels show the results with NS, SS and AS scheme 
respectively.
The sampling resolution increases from left to right in the plots. 
A sub-sampling factor of $f_{\rm
  sub-sample}=0.25,0.025,0.0025$ and $0.00025$ were used 
which resulted in sample sizes of $1.5 \times 10^4$, $1.5 \times 10^5$
,$1.5 \times 10^6$ and $1.5 \times 10^7$ respectively. 
The fraction of stars,
that suffer from scattering in the default SS
scheme, is $3,25,70$ and $93$\% for sample sizes ranging from $1.5 \times 10^4$
to  $1.5 \times 10^7$ (bound systems, which are anyway easily detectable,  
were excluded). 
Hence, for sample sizes below $10^5$ very few 
stars need to be scattered. This fact is also reflected in  
the maps in \fig{radial_velocity}, where,  at resolutions 
below $1.5 \times 10^5$ the maps of NS and SS scheme are 
almost identical. In contrast, the AS scheme significantly 
erodes the structures. 
At resolution of $10^6$, the SS scheme is better than AS but when compared 
to the NS scheme, structures below $r<50 \kpc$ are eroded. 
At even higher resolution,  
the SS scheme is identical to AS scheme, which is expected as
 93\% of the stars are scattered. At the highest resolution,   
most prominent structures are correctly reproduced but the contrast 
of the fine scale structures is reduced. 
Note, these results are without observational 
errors, which when included will anyway erode the fine scale, low 
contrast structures.   
To summarize, we think the results from {\sl Galaxia} when sampling the
stellar halo in 6d space have excellent reliability for resolutions $N<10^5$, 
and for $N=10^6$, moderate reliability only beyond $50 \kpc$. 
In other regimes slight smoothing 
can be expected but the results will still be reliable wherever the 
observational errors are larger than the smoothing effects induced 
by our star spawning process.

We now discuss the nature of our scattering errors.
The uncertainties that arise from our N-body sampling scheme are very 
different from normal observational errors. 
To begin with, our sampling scheme is more like an 
interpolation of a function. This means that if the function 
(density field) is smooth the errors are small. Next, if the number of 
points over which the function is known  is large 
(numerical resolution of the simulations), the errors will again be small.
Finally, the amount of 
scattering in our scheme is inversely proportional to the 
number density of particles, this means that the scattering effects are 
inherently adaptive in space and this is one of the strengths 
of our scheme. For example,  the detectability of a structure
depends upon its density peak being correctly resolved in space 
and this being a high density region has lower amount of 
scattering associated with it. 
Incidentally, the adaptive nature of scattering can also be 
a weakness in some cases. 
For example, if a low particle density region of 
another system contributes significantly to 
the number density of stars at the location of a given structure 
then it can decrease the contrast of that structure. 
This is typically the case in our simulations, where the 
smooth background component of the halo is dominated by a 
few massive accretion events which are not adequately resolved. 
Hence the weakness is not really a drawback of the scheme but more 
an artifact of inadequate particle resolution of simulations.
The question as to what is the appropriate N-body resolution for 
simulating a structure so that it can be properly sampled by our 
scheme, is something which should be tested in future by numerical 
convergence studies of simulations with different resolutions.

Finally, we discuss as to which type of accreting systems are most 
likely to suffer from scattering. 
The BJ05 simulations sample all satellites with typically same 
number of stellar particles (approximately $2 \times 10^4$) 
irrespective of their luminosity, this means low luminosity systems  
are less likely to spawn multiple stars and suffer from 
scattering as compared to the high luminosity ones. 
Moreover, the low luminosity systems are less   
phase mixed and have high particle phase space density 
( as they have smaller internal velocity dispersions and 
are less affected by dynamical friction), and hence these systems 
will also have lower amount of scattering. Finally, dynamical friction 
sinks the systems inwards and it being stronger 
for the high luminosity systems, such systems will lie
preferentially towards the inner regions of the halo than the outskirts. 
This explains the smoothing in the plots being more prominent in the 
inner regions ($r<50 \kpc$).

\section{Discussion}
Presently, given a model  the code generates a particular 
realization of it. While this
has its own applications, e.g., testing instrument capabilities,  
correcting for systematics and making predictions for observations.
But in the end one desires to learn more about the formation of our galaxy. 
And this would require refining or fine tuning the model as 
new data comes in.   
In this respect, our parameterized analytical models based on 
well understood physical 
principles easily lend themselves to model fitting procedures 
and are an invaluable tool for understanding the formation of our 
galaxy. Such schemes have  been successfully used to constrain 
the star formation rate and the age-velocity relation by using the 
data in the solar neighborhood 
\citep{2009MNRAS.397.1286A,2010MNRAS.402..461J}.  A fast and efficient 
method to generate surveys from parameterized models as presented here 
should be immensely useful for performing such 
model fitting analysis in future. More generally one could use the 
Markov Chain Monte Carlo schemes \citep{2002PhRvD..66j3511L} 
to explore the parameter space of the analytical models.  

In order to facilitate the exploration of the parameter space 
there are two issues that need to be addressed.
First, although one can vary the SFR and other model parameters 
that exist in {\sl Galaxia}, dynamical consistency is not guaranteed.  
This can be done by using the scheme outlined 
in \cite{1987A&A...180...94B} (see also
\citet{2010MNRAS.402..461J}), where they self-consistently calculate
the local scale height (or alternatively the ellipticity) as a function of
age. Note, in the above scheme the large scale dynamical self consistency 
is still not guaranteed.
Secondly, although {\sl Galaxia} is fast enough to generate multiple 
realizations of a model but each time the model is changed the code 
needs to recompute the nodes which takes some time. This is also 
something that can be improved. 

Chemical evolution is also another thing that is presently not 
self consistent. Recently some promising semi-analytic schemes 
have been proposed in which chemical evolution is done along 
with radial mixing \citep{2009MNRAS.396..203S}. 
Although presently in {\sl Galaxia} we do not include models 
with radial mixing but we have shown that in principle 
it is possible to accommodate this within our framework.  

There are also some other improvements that can be done.
{\sl Galaxia} has yet to incorporate
a detailed central bar (presently only a bar shaped bulge is included) 
and the spiral arms because these await better 
constraints from
the AAOmega survey of the central disc (K.C. Freeman, personal communication).
Although presently in {\sl Galaxia}, the thick disc is treated as 
a population of constant age
in the manner of the Besan\c{c}on model. This is in agreement with 
\citet{2002ARAA..40..487F}, who consider the thick disc to be
chemically distinct \citep{2003A&A...410..527B,2005A&A...433..185B} 
and uniformly old \citep{1995AJ....109.1095G}.
However, there are alternative models, e.g.,
\citet{2009MNRAS.396..203S} model the thick disc 
by scattering stars from the inner regions of the disc. 
Here the thick disc is also old but not necessarily 
of a constant age. 

Finally, as an application we plan to use {\sl Galaxia}
to address one of the key goals of the HERMES project, i.e., 
recovering star clusters that have long since dispersed within
the Galaxy. The ability to
recover ancient star clusters for a given set of observational
parameters can be tested using a
multi-dimensional group finding algorithm \citet{2009ApJ...703.1061S}.
The power of this approach was recently demonstrated in the context of simulated
dwarf galaxies \citep{2010ApJ...721..582B}. 
With {\sl Galaxia} for a given observational survey one can estimate 
the mass in different galactic components. Subsequently,   
simple prescriptions assuming an initial cluster mass function for
stars along with the assumption of chemical homogeneity with in a cluster  
could be employed to simulate structures in the
chemical abundance space.

\section{Conclusions}
It is only in recent times that the community has begun to undertake
massive spectroscopic surveys of 10$^5$ to 10$^6$ stars. 
These include the RAVE survey on the UK Schmidt 1.3m
telescope \citep{2006AJ....132.1645S}, the APOGEE
\citep{2010IAUS..265..480M} and SEGUE surveys
\citep{2009AJ....137.4377Y} on the Sloan 2.3m telescope  and the upcoming
HERMES survey on the AAT 3.9m telescope \citep{2008ASPC..399..439F}. 
Similar surveys are under discussion for the LAMOST and the
and the CFHT (3.6m) telescopes. These surveys are to be complemented by
huge astrometric missions, in particular, JASMINE
\citep{2008IAUS..248..296Y} and GAIA
\citep{perryman02}, and all-sky
photometric surveys from SkyMapper \citep{2007PASA...24....1K},
PanSTARRS and LSST \citep{2009AAS...21346003I}. 
Thus, within a decade, we will have a vast wealth of data over a 
significant fraction of the Galaxy.

If we are to move beyond information to knowledge, a working framework
will be essential. This framework needs to be sufficiently flexible that it
can adapt to new developments in the theory of galaxy dynamics, updates
in stellar synthesis libraries, and ever-improving N-body
or semi-analytic simulations. On the other hand, the framework should 
also be fast enough 
to generate huge amounts of data and multiple realizations in preparation
for upcoming surveys. To this end, we have presented here a population 
synthesis code for generating a synthetic catalog of stars 
from a given analytical or N body model of a galaxy. 
A set of theoretical isochrones 
are used to convert the given model into a catalog of stars.

Although schemes for converting an analytical model into stellar
catalogs  have been implemented in the past, but they give 
catalogs along specific lines of sight rather than a field of finite
size and its too cumbersome to build a wide field by decomposing 
it into many line of sights.
This makes them limited in their use for generating large 
wide area catalogs. To overcome this limitation, we implement a new algorithm 
for sampling the stars that on one hand generates stars 
that are smoothly distribution over the desired space and on the 
other hand, is very  efficient at generating wide area surveys 
consisting of a large number of stars. 
As a concrete example we have implemented the Besan\c{c}on 
model within {\sl Galaxia} and shown that it can accurately reproduce 
Besan\c{c}on star counts.
However, the design of {\sl Galaxia}  is flexible enough to 
simulate other alternative models also. {\sl Galaxia} in general 
can accept an input model in the form of 
analytic functions for the density distribution, SFR , AMR and AVR.

In addition to sampling an analytical model, 
an algorithm for sampling an N body model is also
presented. The novel feature of this algorithm is its ability 
to sample the stars in accordance with the underlying phase space 
density of the N-body particles. 
As an application we make use of the 
simulated N-body models of the stellar halo by
\citet{2005ApJ...635..931B} which allows one to make detailed 
predictions about the structures in the stellar halo. 
Other than this, the N-body sampling scheme will also be 
useful to simulate known structures in the Milky Way, e.g.,
the streams of the Sagittarius dwarf galaxy. 
Although the N-body sampling scheme is quite robust 
but if the N-body resolution of the simulations is not 
adequate then this can lead to spurious smoothing of the 
structures being sampled.

Using the N-body sampling scheme, we simulated a hypothetical 
GAIA catalog of stars within $4 \kpc$ of the sun and 
analyzed the prospect  of identifying structures in the energy 
and angular momentum space , as in \citet{2010MNRAS.tmp.1196G}.
We find that due to limited numerical resolution of the simulated 
stellar halos, our sampling scheme leads to significant wash out of 
structures. 
However, even if the sampling scheme is not used, 
the distribution of the N-body particles of satellites 
in BJ05 simulations is less clustered as compared to results 
reported by \cite{2010MNRAS.tmp.1196G}. This is something 
that should be investigated in future. 

Although, it is difficult to make very accurate predictions in the solar 
neighborhood, but we show that the BJ05 simulations are suitable for 
analyzing deep large scale surveys of the stellar halo. 
Our analysis shows that the results from {\sl Galaxia} when sampling the
stellar halo in 6d space are very reliable for sample sizes of $N<10^5$, 
e.g., a survey of RR Lyrae or BHB stars.
 In other regimes, slight smoothing 
can be expected but the results will still be reliable whenever the 
observational errors are larger than the smoothing effects induced 
by our star spawning process. 
Additionally, when working in 3d position space the smoothing 
effects can be reduced by calculating the smoothing lengths in 
3d position space only. Our results show that in 3d position space 
most of the structures are accurately reproduced irrespective of 
sample sizes.

In conclusion, we have presented a detailed framework that will allow
present and future stellar surveys to be compared and evaluated 
consistently. To facilitate its wider use, we plan to 
release the {\sl Galaxia} code 
publicly at {\tt http://galaxia.sourceforge.net}.

\section*{Acknowledgments}
JBH is funded through a Federation Fellowship from the Australian 
Research Council (ARC). SS is funded through ARC DP grant 0988751
which supports the HERMES project.
JBH also acknowledges the kind hospitality of Merton College, Oxford,
and a visiting professorship from the Leverhulme Foundation during the 
Hilary and Trinity terms 2010.

\appendix
\section{Rotational kinematics of the disc}
\label{sec:rotational_kin}
The simplest scheme to describe the azimuthal velocity distribution 
is to consider it in the form of a  Gaussian  
along with a term to correct for the asymmetric drift (see equation \ref{equ:veldist}).
A more thorough treatment is obtained by assuming the
\cite{1969ApJ...158..505S} distribution function
\be
f_{\rm Shu}(E,L)  & = &\frac{\gamma(L)\Sigma(L)}{2 \pi \sigma_R^2(L)}{\rm exp} 
\left( \frac{E_c(L)-E}{\sigma_R^2(L)}\right) 
\ee 
where $E_{\rm c}$ is the energy of a circular orbit with angular
momentum $L$  and 
\be
\gamma^2(L)=2/\left(1+\frac{d\ln v_c}{d\ln R_L}\right),
\ee
$R_L=R_c(L)$ being the radius of the circular orbit.

The rms radial dispersion and the surface density is then given by 
\be
\sigma'^{2}_R (R) & = & \frac{\sqrt(2\pi)}{R \Sigma(R)} \int
\frac{\gamma(L)\Sigma(L)}{2 \pi}\sigma_R(L) \times \nonumber \\
& & {\rm exp}\left(\frac{\Phi_{\rm c}(L)-\Phi_{\rm
      eff}(R,L)}{\sigma_R^2(L)}\right) dL
\ee

\be
\Sigma' (R) & = & \frac{\sqrt(2\pi)}{R} \int
\frac{\gamma(L)\Sigma(L)}{2 \pi \sigma_R(L)}  \times \nonumber \\ 
& & {\rm exp}\left(\frac{\Phi_{\rm c}(L)-\Phi_{\rm
      eff}(R,L)}{\sigma_R^2(L)}\right) dL.
\ee
\citep{2009MNRAS.396..203S}. 
Starting with values of  $\sigma_R(L)=\sigma'_R(L)$ and
$\Sigma(L)=\Sigma'(L)$, (primed quantities being target distributions
supplied a priori) one can solve iteratively for the 
values of $\Sigma(L)$ and $\sigma_{R}(L)$ 
using the formalism of \cite{1999AJ....118.1201D}.
The distribution of $L$ and hence $v_{\phi}=L/R$, at a given point $R$
is then given by
\be
P(L,R) & = & \frac{\gamma(R_L)\Sigma(R_L)}{2 \pi \sigma_R(R_L)}{\rm exp}\left(\frac{\Phi_{\rm c}(L)-\Phi_{\rm eff}(R,L)}{\sigma_R^2(R_L)}\right)
\ee
where $\Phi_{\rm c}(L)=\Phi(R_L)+L^2/2R_L^2$ and $\Phi_{\rm
  eff}(R,L)=\Phi(R)+L^2/2R^2$ \citep{2009MNRAS.396..203S}. For a given rotation curve the
potential is computed as $\Phi(R)=-\int_R ^{\infty} v_{\rm c}^2(R) {\rm d}R/R$.
For simplicity we assume a flat rotation curve with $v_c=226.84$. 

Note, $\sigma_R$ is also a function of age. Hence, in practise,  
we calculate the functions  $\sigma_R(L)$ and $\Sigma(L)$ 
for a finite set of ages, and then use the simple nearest neighbor 
interpolation to calculate the $P(L,R)$ for any given age.

\providecommand{\newblock}{} 
\bibliographystyle{apj} 
\bibliography{/home/sharma/texmf/mybib}
\end{document}